\documentclass[amsmath, amssymb, twocolumn, aps, prx, superscriptaddress, footinbib]{revtex4-1}

\usepackage{mathtools}
\usepackage{mathrsfs, amsmath, wasysym}
\usepackage{bbold}
\usepackage[mathscr]{eucal}
\usepackage{xfrac}
\usepackage{graphicx}
\usepackage{dcolumn}
\usepackage{bm}
\usepackage{color}
\usepackage{tabularx}
\usepackage{natbib}
\usepackage[colorlinks,linkcolor=blue,citecolor=blue,urlcolor=blue]{hyperref}
\usepackage[normalem]{ulem}
\usepackage[all]{hypcap}
\usepackage[dvipsnames,usenames,table]{xcolor}
\usepackage{braket,mleftright}

\newcommand{\nocontentsline}[3]{}
\newcommand{\tocless}[2]{\bgroup\let\addcontentsline=\nocontentsline#1{#2}\egroup}

\newcommand{\bz}{{\bf z}}

\newcommand{\bs}{{\bf s}}
\newcommand{\bS}{{\bf S}}

\newcommand{\bw}{{\bf w}}

\newcommand{\bn}{{\bf n}}

\newcommand{\bT}{{\bf T}}

\newcommand{\bsigma}{\boldsymbol{\sigma} }

\newcommand{\blambda}{\boldsymbol{\lambda} }

\newcommand{\ehat}{\hat {\bf e}}

\newcommand{\be}{\begin{equation}}
\newcommand{\ee}{\end{equation}}
\newcommand{\beg}{\begin{gather}}
\newcommand{\eeg}{\end{gather}}
\newcommand{\beq}{\begin{eqnarray}}
\newcommand{\eeq}{\end{eqnarray}}
\newcommand{\bea}{\begin{align}}
\newcommand{\eea}{\end{align}}
\newcommand{\beqq}{\begin{eqnarray*}}
\newcommand{\eeqq}{\end{eqnarray*}}

\newcommand{\tr}[1]{\text{Tr}(#1 )  }

\newcommand{\up}{\uparrow}
\newcommand{\down}{\downarrow}

\begin{document}


\title{Canonical strong coupling spin wave expansion of Kondo lattice magnets. \\
I. Effective Hamiltonian via canonical transformation}

\author{J. Strockoz}
\thanks{These authors contributed equally to this work.}
\affiliation{Department of Physics, Drexel University, Philadelphia, PA 19104, USA}%

\author{M. Frakulla}
\thanks{These authors contributed equally to this work.}
\affiliation{Department of Physics, Drexel University, Philadelphia, PA 19104, USA}%

\author{D. Antonenko}
\affiliation{Department of Physics, Drexel University, Philadelphia, PA 19104, USA}%
\affiliation{Department of Physics, Yale University, New Haven, CT 06520, USA}%

\author{J. W. F. Venderbos}
\affiliation{Department of Physics, Drexel University, Philadelphia, PA 19104, USA}%
\affiliation{Department of Materials Science \& Engineering, Drexel University, Philadelphia, PA 19104, USA}%

\begin{abstract}
This paper develops a systematic strong coupling spin wave expansion of itinerant Kondo lattice magnets, magnets in which local moment spins are Kondo coupled to itinerant charge degrees of freedom. The strong coupling expansion is based on a canonical Schrieffer-Wolff transformation of the Hamiltonian, which is performed after $1/S$ expansion of the local moments and determined iteratively by requiring that spin-flip terms are removed at each order.  We demonstrate that the canonical transformation can be viewed as an order-by-order diagonalization of the quantum Kondo coupling---the dominant term in the strong coupling regime. A consequence is that the transformed electron operators correspond to electrons in a state of total spin $S\pm 1/2$ with the local moments, and the transformed boson operators describe spin wave excitations of the total local spin. We show that the electron degrees of freedom can be thought of as tightly bound spin polarons. We further show that the strong coupling spin wave expansion is readily extended to include the effects of spin-orbit coupling or electron pairing. 
\end{abstract}

\date{\today}


\maketitle



\section{Introduction \label{sec:intro}}

Within the realm of condensed matter physics, a broad class of physical phenomena are well-described by a relatively simple model of local magnetic moments interacting with itinerant charge carriers. Hallmark examples of such phenomena include the Ruderman-Kittel-Kasuya-Yosida (RKKY) interaction between magnetic impurities in metals~\cite{Ruderman:1954p99,Kasuya:1956p45,Yosida:1957p893}, double-exchange magnetism~\cite{Zener:1951p440,Anderson:1955p675,Dagotto:2001p1,Izyumov:2001p109}, ferromagnetism in diluted magnetic semiconductors~\cite{Konig:2000p5628}, as well as Kondo physics in heavy fermion materials. More recently, models based on spin-charge interactions have attracted considerable attention as possible venues for realizing various topological electronic phases. These include topological Chern bands generated by magnetic textures with real space topology~\cite{Ye:1999p3737,Onoda:2004p2624,Hayami:2014p085124,Hamamoto:2015p115417}, Chern insulating phases stabilized by noncoplanar magnetic orders~\cite{Ohgushi:2000pR6065,Martin:2008p156402,Akagi:2010p083711,Kumar:2010p216405} and, perhaps most notably, topological superconductivity induced by depositing arrays of magnetic atoms on conventional superconductors (or schemes similar to this)~\cite{Choy:2011p195442,Martin:2012p144505,Nadj-Perge:2013p020407,Klinovaja:2013p186805,Vazifeh:2013p206802,Braunecker:2013p147202}. Signatures of the latter have been observed in experiment~\cite{Nadj-Perge:2014p602,Yazdani:2015p014012}. 

Despite the fact that these phenomena have a simple common root---local moment spins coupled to itinerant electrons---the physics can nonetheless differ vastly depending on specific characteristics such as the strength of the spin-charge exchange coupling relative to other energy scales, the sign of the spin-charge coupling (i.e., ferromagnetic versus antiferromagnetic), and the concentrations of charge carriers and local moments. 

In many specific models of spin-charge coupling, in particular those aimed at studying magnetism, the local moments are treated as classical spins. A prominent example is the Kondo lattice model, which describes itinerant charge carriers coupled to a regular lattice of spins. The Kondo lattice model has been demonstrated to provide a good description of a variety of itinerant magnetic systems~\cite{Dagotto:2001p1,Izyumov:2001p109}, of which the aforementioned magnetic superconductor hybrid structures are but one example. Treating the local moments as quantum spins in general gives rise to a complicated  Hamiltonian of interacting electrons and quantum spins, which is tractable only in certain limiting cases~\cite{Anderson:1955p675,Kubo:1972p21,Kubo:1972p929,Muller-Hartmann:1996pR6819,Shen:2000p9532}.  

One approach to going beyond a classical treatment is expanding around the classical limit in $1/S$, where $S $ is the length of the spins. That is the approach we take in this paper. In particular, we develop a systematic $1/S$ spin wave expansion based on a canonical Schrieffer-Wolff transformation of the Kondo lattice Hamiltonian. 

The description of spin wave excitations in itinerant Kondo lattice magnets comes with complications that do not arise in local moment magnets of the Heisenberg type. In particular, in Kondo lattice magnets fluctuations of the local moment spins couple to the charge carriers, thus giving rise to a theory of bosons interacting with electrons~\cite{Kubo:1972p21,Furukawa:1996p1174}. Furthermore, in cases where such coupling is large, it would physically be more natural to treat fluctuations of the combined total spin as the fundamental spin wave excitations~\cite{Shannon:2002p104418}. A systematic theory for fluctuations of the total local moments is not straightforwardly obtained from the degrees of freedom of the original Kondo lattice model, however, in particular since the total spin at a given site is not fixed due to electron itinerancy. The spin wave expansion scheme we present here confronts and overcomes the challenges commonly encountered when addressing spin wave excitations in itinerant Kondo lattice magnets.

The spin wave expansion introduced in this paper is a strong coupling expansion. It assumes that the Kondo coupling $J_K$, i.e., the local exchange coupling between the local moments and the itinerant electrons, is the largest energy scale. Its construction proceeds in two steps. First, the local moment spins are expressed in terms of Holstein-Primakoff bosons---this is the standard starting point of $1/S$ expansions. This first step is followed by the second and most important step, which is a strong coupling expansion via unitary canonical transformation of the Hamiltonian. The canonical transformation is designed to remove electron spin-flip processes to any desired order in $1/S$ as well as $t/J_K$, where $t$ is the energy scale associated with electron hopping. By spin-flip terms we refer to those terms which flip the spin of an electron from locally aligned to locally anti-aligned with the classical moment, or vice versa. Such terms change the relative number of electrons with aligned and anti-aligned spins and therefore connect sectors of the Hilbert state separated by the large energy scale $J_K$ in the strong coupling limit. Eliminating spin-flip terms is thus provides a way to integrate out couplings between these low- and high-energy sectors, which is the usual goal of strong coupling expansions~\cite{MacDonald:1988p9753,Chernyshev:2004p235111}.

In the present case, we will show that, after canonical transformation, the Kondo lattice Hamiltonian is expressed in terms of the natural low-energy degrees of freedom of the strong coupling regime~\cite{Shannon:2002p104418}. That is to say, the transformed bosons correspond to spin wave excitations of the total spin $ \bS_i + \bs_i$ (local moment plus electron spin) at a given site $i$ and the transformed fermion operators create (or annihilate) electrons in a state of total spin $S\pm 1/2$, i.e., an eigenstate of $ \bS_i + \bs_i$ rather than $\bs_i$. This property constitutes the key appeal of the spin wave expansion developed in this work.

After projecting out the high energy fermions the strong coupling expansion results in a Hamiltonian for effectively spinless fermions coupled to bosonic spin wave excitations. Quantum effects of the local moments are systematically accounted for to any desired order in $1/S$ and couplings to the high energy fermions are accounted for perturbatively in $t/J_K$. The effective Hamiltonian, which requires only a single fermion flavor, provides a full description of the low-energy dynamics of itinerant Kondo lattice magnets and the fact that it effectively only requires a single fermion flavor is yet another appealing feature of the spin wave expansion. In light of these key characteristics, which will be further elucidated in the remainder of this paper, as well as the fact that the spin wave expansion crucially relies on a canonical transformation, we refer to it as a canonical (strong coupling) spin expansion for itinerant magnets. 

Previous studies of spin wave excitations in Kondo lattice magnets have largely focused on itinerant ferromagnets~\cite{Kubo:1972p21,Furukawa:1996p1174,Nagaev:1998p827,Golosev:2000p3974,Perkins:1999p1182,Shannon:2002p104418}, with only very few exceptions~\cite{Lv:2010p045125,Akagi:2013p123709,Akagi:2014p014017}. The canonical spin wave theory developed in this work makes no particular assumptions regarding the ordered ground state and by design applies to all types of classical spin configurations. This generality allows for a straightforward expansion around noncollinear or noncoplanar spin states. Spin states of this kind can arise from the competition of the Kondo coupling, which tends to favor ferromagnetism, in particular when it is large, and the Heisenberg coupling between the local moments. Our theory therefore includes an additional generic Heisenberg exchange coupling from the outset. This has the consequence that the spin waves acquire their dispersion both directly from the Heisenberg term and the coupling to the fermions via self-energy effects. 

The canonical spin wave expansion can be extended in a straightforward manner to systems in which spin-orbit coupling is important. Spin-orbit coupling affects both the itinerant electrons, by locking spin to orbital motion, and the magnetic exchange interactions between the local moments by giving rise to anisotropic exchange interactions~\cite{Banerjee:2014p031045,Meza:2014p085107,Hayami:2018p137202,Okada:2018p224406,Zhang:2020p024420,Kathyat:2020p075106,Kathyat:2021p035111}. We will show that accounting for both effects can be achieved with only minor modifications, thus considerably expanding the direct applicability of the canonical spin wave theory. 

A second and more consequential generalization is achieved by including a superconducting pairing term of the itinerant electrons, such that the generalized model describes magnetic superconductors. Magnetic superconductors can arise due to the coexistence of intrinsic superconductivity and magnetism, which in such cases typically exhibits spatial modulation, or can be realized by inducing pairing via the proximity effect. Engineered hybrid systems of the latter kind have attracted much attention as a possible platform for realizing topological superconductivity~\cite{Choy:2011p195442,Martin:2012p144505,Nadj-Perge:2013p020407,Klinovaja:2013p186805,Vazifeh:2013p206802,Braunecker:2013p147202}. We will demonstrate that the canonical spin wave expansion is readily generalized to magnetic superconductors with conventional $s$-wave pairing, and will show that the spin wave excitations directly couple to the pairing. The pairing therefore affects the spin wave dispersion, and the coupling to the spin waves may also affect the pairing. 

This paper focuses on the formal development of the strong coupling spin wave expansion and its generalizations. In a companion paper we apply the spin wave expansion to a number of specific lattice models. The goal of these applications is to study various properties of spin wave excitations in Kondo lattice magnets which the spin wave expansion allows to address.


\section{Itinerant Kondo lattice magnets  \label{sec:klm}}

We begin by introducing the Kondo lattice model, which describes a broad class of itinerant magnets. The Kondo lattice Hamiltonian takes the form
\begin{multline}
H = \sum_{ij} t_{ij} c^\dagger_i c_j -  \frac{J_K}{2S}\sum_i \bS_i \cdot c^\dagger_i \bsigma c_i \\ + \frac{1}{2S^2}\sum_{ij} J_{ij} \bS_i\cdot \bS_j,  \label{eq:H_klm}
\end{multline}
where $c^\dagger_{i} = (c^\dagger_{i\up}, c^\dagger_{i\down})$ are the electron operators which create electrons at site $i$ with spin $\sigma=\up,\down$, and $\bS_i$ are the local moment spins with spin quantum number $S$. To describe the itinerant electron spin we have made use of a set of a Pauli spin matrices $\bsigma = (\sigma_x,\sigma_y,\sigma_z)$. The first term in the Hamiltonian describes the hopping of electrons between different sites $i,j$ with hopping amplitudes $t_{ij}$ and thus corresponds to a kinetic term for the itinerant charge carriers. We assume that the hopping amplitudes are real and satisfy $t_{ij} = t_{ji}$. (The more general case of spin-dependent hoppings stemming from spin-orbit coupling is considered in Sec.~\ref{sec:soc}.) The second term describes the Kondo coupling between the local moment spins and the itinerant electrons with coupling constant $J_K$. Note that for later convenience we have explicitly separated out the length $S$ of the spins. 

The third term describes a direct Heisenberg exchange coupling between the spins with exchange coupling constants $J_{ij}=J_{ji}$; the factor $1/2$ is introduced to avoid double counting. Including the direct Heisenberg exchange coupling makes the Hamiltonian of Eq.~\eqref{eq:H_klm} more general than what is often understood as the Kondo lattice Hamiltonian. In what follows we will nonetheless refer to Eq.~\eqref{eq:H_klm} as the Kondo lattice model, but emphasize here that we mean the more general variant which generically includes a Heisenberg coupling.

Since this paper is specifically focused on the regime of strong Kondo coupling, in which $J_K$ is assumed to be largest energy scale, it is useful to first consider the Kondo coupling on a single site---the Kondo ``atom''. This will also provide a basic understanding of the physics described by the Kondo lattice Hamiltonian. The Kondo coupling at a single site $i$ can be written as   
\be
H_{K,i} =-\frac{J_K}{2S} \bS_i  \cdot c^\dagger_i \bsigma c_i = -\frac{J_K}{S} \bS_i \cdot \bs_{ i }, \label{eq:H_Ki}
\ee
where $\bs_i = c^\dagger_i \bsigma c_i /2$ is the electron spin. It is straightforward to diagonalize \eqref{eq:H_Ki} in terms of the total spin $\bT_i = \bS_i + \bs_{i }$ at site $i$. If site $i$ is unoccupied by an electron, or is doubly occupied, the energy eigenvalue is simply $\varepsilon_0 = \varepsilon_2 = 0$. Instead, if site $i$ is singly occupied and the two spins form a state with total spin $T=S\pm 1/2$, the eigenvalues of the Kondo term are
\be
\varepsilon_+ = -J_K/2, \qquad \varepsilon_- = J_K(S+1)/2S, \label{eq:eps_pm}
\ee 
where $\varepsilon_\pm $ corresponds to $T=S\pm 1/2$. Using the quantum theory of angular momentum it is straightforward to write down the eigenstates $\ket{S\pm \tfrac12,M}$ corresponding to the solutions $\varepsilon_\pm $. The eigenstates are given by 
\beq
\ket{S+\tfrac12,M} &=& u \ket{S,M-\tfrac12; \up} + v \ket{S,M+\tfrac12; \down},  \label{eq:psi_+} \\
\ket{S-\tfrac12,M} &=& -v \ket{S,M-\tfrac12; \up} + u \ket{S,M+\tfrac12; \down},  \label{eq:psi_-}
\eeq
where $u=u_M$ and $v=v_M$ are the appropriate Clebsch-Gordan coefficients (which depend on the magnetic quantum number $M$) and are given by 
\be
u_M = \sqrt{\frac{S+1/2+M}{2S+1}}, \quad v_M = \sqrt{\frac{S+1/2-M}{2S+1}}. \label{eq:u,v}
\ee

An immediate and important observation is that the quantum Kondo atom is quite different from the classical Kondo atom. In the classical problem the spins are treated as classical vectors and the problem effectively reduces to an electron spin in a magnetic field along $\bS_i$. The energies become $\varepsilon_\pm \rightarrow \mp J_K/2$ and the eigenstates correspond to spins aligned or anti-aligned with the classical moment. The energies are clearly symmetric with respect to changing the sign of the Kondo coupling, $J_K \rightarrow -J_K$, and this reflects a well-known local gauge-like symmetry of the classical Kondo lattice Hamiltonian: the sign of the Kondo coupling can be changed by a local unitary rotation of the electron spin quantization axis and therefore does not affect the physics. What is referred to as ``aligned'' or ``anti-aligned'' depends on the choice of quantization axis.

The quantum problem is clearly different. The energies are \emph{not} symmetric with respect to the change $J_K \rightarrow -J_K$ and neither is the structure of the eigenstates. There are $2S$ states with eigenvalues $\varepsilon_-$ and $2S+2$ states with eigenvalues $\varepsilon_+$. Furthermore, the states $\ket{S-\tfrac12,M} $ are never product states, whereas $\ket{S+\tfrac12,S+\tfrac12} =\ket{S,S; \up} $ is a product state. The latter explains why the eigenvalue $\varepsilon_+$ in \eqref{eq:eps_pm} coincides with one of the classical energies. More fundamentally, in the quantum case there is no notion of aligned and anti-aligned spins. Quantum mechanically it only makes sense to speak of states with total spin $S\pm 1/2$.   

It follows from this observation that if a strong coupling spin wave expansion of the Kondo lattice model is to accurately capture the quantum nature of the spins, at least to a given order in $1/S$, it should be sensitive to the sign of $J_K$, and recover the distinct nature of the solutions of the quantum Kondo atom. As will be shown, the proposed canonical spin wave expansion indeed satisfies this criterion, and when applied to the Kondo atom problem indeed reproduces the exact energy eigenvalues of Eq.~\eqref{eq:eps_pm}. A detailed discussion of this, and of connections to previous expansions, is presented in Sec.~\ref{ssec:previous-work}.

As mentioned in the introduction, while the Kondo lattice model captures a variety of different correlated phenomena, in this paper we focus on the case where the ground state of the Kondo lattice model is magnetically ordered. A large body of literature has been devoted to studying the phase diagram of the Kondo lattice model, in particular as far as magnetism is concerned. It is not our intention here to review this literature at any level of detail, but a few general remarks may be nonetheless be helpful. 

The first remark concerns the two opposing limits of the Kondo coupling: the weak and strong coupling limits. The weak coupling limit, in which $J_K$ is much smaller than the electron band width, is generally referred to as the RKKY limit, and in this limit the problem of magnetism can be cast in terms of an effective interaction between the local moments, mediated by the itinerant electrons. The well-known RKKY interaction is long-ranged and has an oscillatory tail. The opposite limit (i.e., the strong coupling limit) is generally referred to as the double-exchange limit, at least when the spins are treated classically or when the Kondo coupling is ferromagnetic.  Since in this limit the electron spin is tied to the local moment at each site, magnetic ordering becomes a question of electron kinetic energy: the spin configuration which minimizes the electron kinetic energy is favored as the magnetic ground state. This is known as the double-exchange mechanism for magnetic ordering and finds its historical origin in the specific context of the manganese oxides \cite{Zener:1951p403,Anderson:1955p675}. 

The second remark concerns the origin of the sign of the Kondo coupling. An antiferromagnetic Kondo coupling ($J_K<0$ in our convention) is naturally obtained when the Kondo lattice Hamiltonian is derived from the periodic Anderson model using second order perturbation theory~\cite{Schrieffer:1966p491}, in which case $S=1/2$. The periodic Anderson model is a common starting point for describing the physics of heavy fermion systems. A ferromagnetic Kondo coupling, on the other hand, arises naturally in context of strongly correlated electron systems with orbital degrees of freedom. Physically this is due to Hund's rule coupling, and it is therefore common in this context to simply refer to the Kondo coupling as Hund's rule coupling. The canonical spin wave expansion developed in this work does not rely on any assumption regarding the nature of the Kondo coupling, and is therefore broadly applicable.

The third and concluding remark concerns the generic ordering tendency of the Kondo lattice model. The Kondo lattice model without direct Heisenberg exchange interactions is known to have a distinct tendency towards ferromagnetism. In particular, in the simple classical-spin model of a two-site dimer at strong coupling, the spins are ferromagnetically coupled in the presence of a single electron, and antiferromagnetically coupled in the presence of two electrons~\cite{Zener:1951p440,Anderson:1955p675}. There is no coupling when no electrons are present. As such, the two-site model already demonstrates a dependence of the ordering tendency on electron density. An additional (antiferromagnetic) Heisenberg interaction between the local moments competes with this tendency, and this competition can give rise to a rich magnetic phase diagram and is relevant to a wide array of strongly correlated systems.


\section{Canonical spin wave expansion \label{sec:transform}}

This section presents the derivation of the main result of this work: a systematic strong coupling spin wave expansion for Kondo lattice magnets. As is the case for any spin wave expansion, the starting point is a magnetically ordered ground state defined by a configuration of classical spins. In the case of Heisenberg spin models the standard approach to constructing a spin wave expansion is to first choose a local spin coordinate frame such that the $z$-axis is along the direction of the ordered moment at each site, and then perform the Holstein-Primakoff boson substitution. The resulting Hamiltonian for the spin wave bosons is a systematic expansion in $1/S$ and can be used to compute properties of the excitations of the classical ground state. The zeroth order term in this expansion is simply the classical energy of the ground state and the next-order quadratic piece of the boson Hamiltonian forms the basis of linear spin wave theory (i.e., non-interacting bosons).  

\begin{figure}
	\includegraphics[width=\columnwidth]{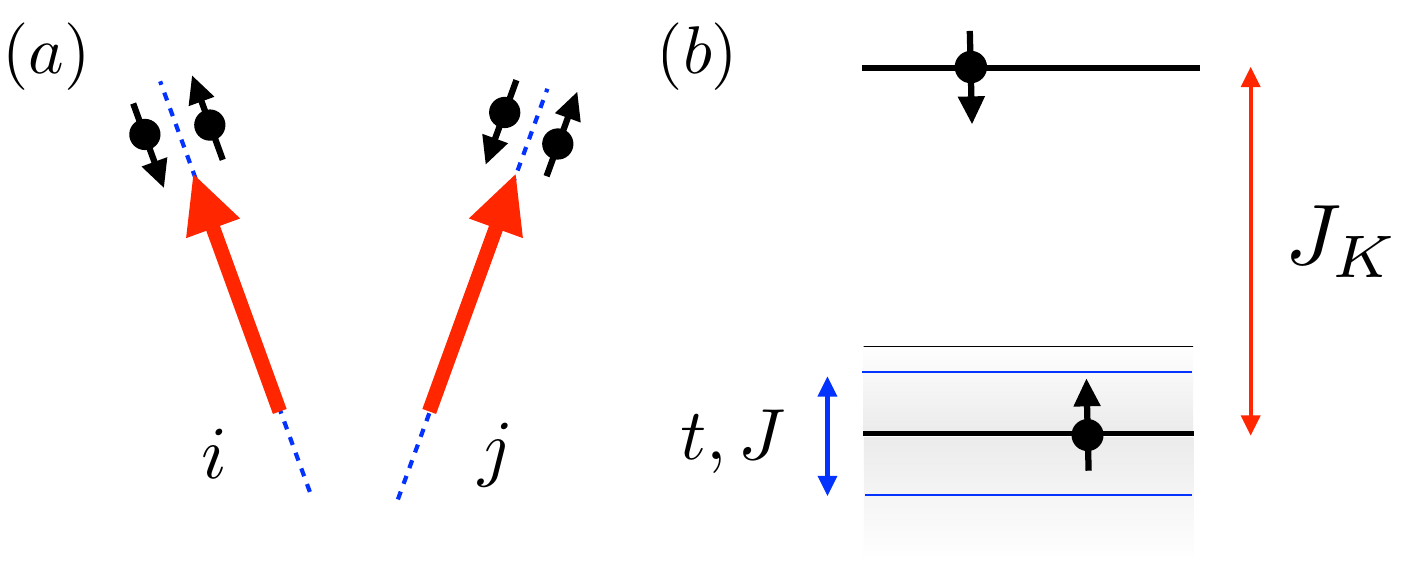}
	\caption{(a) In the spin wave expansion the quantization axis of the electron spin is chosen along the direction of the ordered classical moments at each site. (b) Separation of energy scales underlying the strong coupling spin wave expansion.}
	\label{fig:sketch}
\end{figure}

In the case of the Kondo lattice model, adopting a coordinate frame tied to the direction of the ordered moments also fixes the quantization axis for the electron spin along the classical spins. The quantization axis is therefore in general site-dependent. This has the implication that spin-up and spin-down should be understood as aligned and anti-aligned with the local moment, as shown schematically in Fig.~\ref{fig:sketch}(a). Once the Hamiltonian of Eq.~\eqref{eq:H_klm} is expressed in terms of the local frame spin variables, the Holstein-Primakoff substitution can be performed and this yields a $1/S$ expansion of the Kondo lattice Hamiltonian around its classical limit, i.e., the limit in which the spins are replaced by classical vectors. All terms in this expansion beyond the classical limit correspond to interactions between the spin wave bosons and the electrons, as well as the usual pure boson terms originating from the Heisenberg term in Eq.~\eqref{eq:H_klm}. 

The classical limit of the Kondo lattice model corresponds to non-interacting electrons coupled to a background of classical moments, all polarized in the local $z$ direction. Since the quantization axis is aligned with the local moments (making the Kondo coupling manifestly diagonal) the hopping between sites is generally spin-dependent. When the Kondo coupling is large, i.e., much larger than any other energy scale in the problem, there is a large energetic separation between the local spin-up and spin-down states. This separation of energy scales in the regime of large Kondo coupling is shown schematically in Fig.~\ref{fig:sketch}(b). In this paper we assume that the Kondo coupling is indeed large, such that it is sensible to seek a theory for one of the electron spin species only. Given this assumption, the central aim of this paper is to construct an effective theory for the low-energy degrees of freedom by means of a perturbative strong coupling expansion, which is thus performed after the initial $1/S$ expansion of the local moment spins $\bS_i$. The strong coupling expansion is based on a canonical Schrieffer-Wolff transformation and as such provides an elegant systematic way to obtain an effective Hamiltonian. This resulting effective theory constitutes both a systematic $1/S$ spin wave expansion and an expansion in the small parameters $t/J_K$ and $J/J_K$~\footnote{Here and in what follows we simply write $t/J_K$ and $J/J_K$ to denote the expansion parameters, with the understanding that these should be read as $|t_{ij}|/J_K$ and $|J_{ij}|/J_K$. The expansion allows general hopping and exchange parameters $t_{ij}$ and $J_{ij}$ beyond nearest neighbors.}, as will be shown in detail in the remainder of this section and this paper. 

The derivation of the strong coupling expansion relies on the requirement that terms which change the number of low-energy and high-energy electrons are removed to given order in the expansion parameters. There are two types of such terms: spin-flip hopping processes and spin-flip processes in which a magnon is emitted or absorbed. The latter originate from the spin-flip part of the Kondo coupling. Removing such terms perturbatively via canonical transformation generates a Hamiltonian which acts only within the subspaces defined by a fixed number of spin-up and spin-down electrons. When no high energy states are occupied one may project onto the low-energy subspace and obtain an effectively spinless Hamiltonian, which can be determined to the desired order in $1/S$ and $t/J_K$. 

In the remainder of this section, we describe the derivation of the strong coupling spin wave expansion in detail. The key step in this derivation is the construction of a canonical transformation which removes the spin-flip terms. In Sec.~\ref{ssec:derivation} we  specify an iterative procedure for determining the canonical transformation. In the final part of this section we describe how to assemble the effective Hamiltonian from the formal strong coupling expansion.

\subsection{Local spin basis \label{ssec:local-spin}}

The first step is to introduce a local quantization $\hat \bn_i$ aligned with the ordered moments of the classical ground state. The unit vectors $\hat \bn_i$ define the local $z$ axis and are specified in terms of the angles $\{\theta_i,\phi_i\}$ as
\be
\hat \bn_i =  (\cos\phi_i\sin\theta_i,\sin\phi_i\sin\theta_i,\cos\theta_i). \label{eq:n_i}
\ee
This choice of quantization axis implies new electron operators $f^\dagger_{i\sigma}$ and $f_{i\sigma}$, which create and annihilate electrons with spin $\sigma=\up,\down$ along the local axis, see Fig.~\ref{fig:sketch}(a). These $f$-fermions are related to the $c$-fermions by a local unitary transformation $U_i$ as
\be
 f^\dagger_i =  c^\dagger_i U_i, \quad   f_i =  U^\dagger_i c_i. \label{eq:f-fermions}
\ee
The matrix $U_i$ is given by $U_i =( \bz_i , \bw_i )$,  where
\be
\bz = \begin{pmatrix} \cos \frac{\theta}{2} \\ \sin \frac{\theta}{2}e^{i\phi} \end{pmatrix}, \quad \bw =\begin{pmatrix} - \sin \frac{\theta}{2}e^{-i\phi}\\  \cos \frac{\theta}{2} \end{pmatrix}, \label{eq:z-w}
\ee
are spinors corresponding to an electron state with spin along the local up ($\bz$) and local down ($\bw$) direction. Using these spinors, we can define projectors $P^\pm_i$ on the subspaces corresponding to locally aligned ($+$) and anti-aligned spins ($-$). These projectors are given by 
\be
P^+_i = \bz_i \bz^\dagger_i  , \quad P^-_i = \bw_i \bw^\dagger_i ,    \label{eq:projectors}
\ee
and can be reduced to the form $P^\pm_i  = (1 \pm \hat \bn_i\cdot \bsigma)/2$.

Next, we write the spins $\bS_i$ in terms of the local frame as 
\be
\bS_i = S^x_i \hat{\bf e}^x_i+S^y_i \hat{\bf e}^y_i+S^z_i \hat{\bf e}^z_i , \label{eq:S_i}
\ee
where $(S^x_i,S^y_i,S^z_i)$ denote the spin components with respect to a local orthonormal frame $(\hat{\bf e}^x_i,\hat{\bf e}^y_i,\hat{\bf e}^z_i)$, defined such that $\hat{\bf e}^z_i $ is equal to $\hat \bn_i$ given by Eq.~\eqref{eq:n_i}. The local frame vectors can be collected in a rotation matrix $R_i$,
\be
R_i = \begin{pmatrix} \ehat^x_i & \ehat^y_i & \ehat^z_i \end{pmatrix},
\ee
which relates the spin components of the fixed global frame to the local frame as $\bS_i \rightarrow R_i \bS_i$. 

We are now in a position to express the Hamiltonian of Eq.~\eqref{eq:H_klm} in terms of the $f$-fermions and the spin components of the local frame. This yields
\begin{multline}
H = \sum_{ij} t_{ij} f^\dagger_i \Omega_{ij} f_{j}  - \frac{J_K}{2S}\sum_i  S^a_i f^\dagger_i \sigma^a f_i \\
+ \frac{1}{2S^2}\sum_{ij} J_{ij}\Gamma^{ab}_{ij} S^a_iS^b_j. \label{eq:H_local}
\end{multline}
Here the matrix $\Omega_{ij} $ is given by 
\be
\Omega_{ij} \equiv U^\dagger_i U_j = \begin{pmatrix} \bz^\dagger_i\bz_j & \bz^\dagger_i\bw_j  \\ \bw^\dagger_i\bz_j  & \bw^\dagger_i\bw_j  \end{pmatrix}, \quad \Omega^\dagger_{ij} = \Omega_{ji} ,\label{eq:Omega}
\ee
and reflects the fact that in general the hopping becomes spin-dependent when the spin basis changes from site to site. The overlap $\bz^\dagger_i\bw_j $, for instance, which is overlap of a spin-$\up$ state at site $i$ with a spin-$\down$ state at site $j$, is nonzero unless the classical moments at site $i$ and $j$ are aligned. For further analysis it is convenient to absorb this spin-dependence in the definition of the hopping by defining a hopping tensor
\be
t^{\sigma\sigma'}_{ij} = t_{ij} (\Omega_{ij})^{\sigma\sigma'} , \qquad ( t^{\sigma\sigma'}_{ij})^* = t^{\sigma'\sigma}_{ji}. \label{eq:t_ij_ss} 
\ee

In the third term of Eq.~\eqref{eq:H_local}, the Heisenberg exchange term, the matrix $\Gamma^{ab}_{ij} $ is defined by the overlap of the local frame unit vectors on different sites, i.e., 
\be
  \Gamma^{ab}_{ij}  = \ehat^a_i\cdot \ehat^b_j \label{eq:Gamma},
\ee
and thus captures the way in which the local frames differ from site to site. In matrix form, $\Gamma_{ij}$ can be written as
\be
\Gamma_{ij} = R^T_i R_j, \qquad  \Gamma^T_{ij} = \Gamma_{ji}, \label{eq:Gamma_ij}
\ee
which emphasizes that $\Gamma_{ij}$ is the SO(3) analog of $\Omega_{ij}$ defined in \eqref{eq:Omega}. In preparation for the Holstein-Primakoff substitution, it is useful to rearrange the Heisenberg term in Eq.~\eqref{eq:H_local} (which is denoted $H_J$) to obtain
\begin{multline}
H_J = \frac{1}{2S^2} \sum_{ij} J_{ij} \left\{ \Gamma^{zz}_{ij} S^z_i S^z_j + \left[ \frac12 A_{ij }S^-_iS^+_j + \right.\right. \\
\left.\left.  \frac12 B_{ij }S^-_iS^-_j + (\Gamma^{zx}_{ij} +i\Gamma^{zy}_{ij} )S^z_i S^-_j +\text{H.c.} \right] \right\} \label{eq:H_J}
\end{multline}
with coefficients $A_{ij}$ and $B_{ij}$ are given by
\be
\begin{split}
A_{ij } & =  (\Gamma^{xx}_{ij} + \Gamma^{yy}_{ij}-i \Gamma^{xy}_{ij}+ i\Gamma^{yx}_{ij})/2, \label{} \\
B_{ij } & =   (\Gamma^{xx}_{ij} - \Gamma^{yy}_{ij}+i \Gamma^{xy}_{ij}+i\Gamma^{yx}_{ij})/2. 
\end{split} \label{eq:A_ijB_ij}
\ee
Here we have made use of the property $\Gamma^{ab}_{ij}=\Gamma^{ba}_{ji}$.

\subsection{Holstein-Primakoff bosons \label{ssec:HPboson}}

As is standard in spin wave expansions, the next step is to express the spin operators $\bS_i$ in terms of Holstein-Primakoff (HP) bosons $a_i$~\cite{Holstein:1940p1098,Holstein:1941p388}. Specifically, we write
\be
\begin{split}
S^z_i &= S - a^\dagger_i a_i,  \\
 S_i^+ &=  (2S - a^\dagger_i a_i )^{1/2} a_i ,  \\
 S_i^- &= a^\dagger_i(2S - a^\dagger_i a_i )^{1/2},  
\end{split} \label{eq:HP}
\ee
where $ S_i^\pm  = S_i^x\pm iS_i^y$.  We stress that the HP substitution is performed in the local frame, i.e., the spin components $(S^z_i,S^+_i,S^-_i)$ in \eqref{eq:HP} are with respect to the local frame, as defined in Eq.~\eqref{eq:S_i}. Note that writing spin variables in terms of bosons enlarges the Hilbert space, but the operators defined by \eqref{eq:HP} do not couple the physical and unphysical subspaces. For the HP substitution to be useful, however, the square root is expanded, which may be justified when fluctuations are small (i.e., $\langle n_i \rangle/2S\ll 1$). In this case the strict decoupling of the physical and unphysical subspaces no longer holds and in general the consistency and validity of the HP spin wave expansion must be checked afterwards by computing $\langle n_i \rangle$.

After performing the HP substitution in the Hamiltonian of Eq.~\eqref{eq:H_local}, the resulting Hamiltonian can be conveniently grouped into four distinct terms, which we write as
\be
H = H_0 + T_0 + T_1 + T_{-1}. \label{eq:H_Tm}
\ee
Here the first term $H_0$ is the classical limit of the Kondo coupling (in the local frame) and is given by
\be
H_0 = \sum_{i,\sigma} \varepsilon_\sigma f^\dagger_{i\sigma} f_{i\sigma},   \label{eq:H_0}
\ee
with classical Kondo energies $\varepsilon_\sigma$ given by
\be
\varepsilon_{\up,\down} = \mp J_K/2.
\ee
These are simply the energies of an electron which is aligned ($\varepsilon_{\up} $) or anti-aligned ($\varepsilon_{\down} $) with the local moment. Since $J_K$ is large $H_0$ describes a large splitting between the two spin states, and the sign of $J_K$ determines which spin species forms the low-energy band.

The remaining terms of the Hamiltonian are grouped into three operators $T_m$, with $m=0,\pm1$, such that $T_0$ collects all terms which do not flip the electron spin, and $T_{\pm 1}$ collect all spin flip terms. The latter thus collect all terms which connect the high and low-energy states as defined by \eqref{eq:H_0}. 

More precisely, the operator $T_0$ takes the form
\be
T_0 = T^{(t)}_0+T^{(S)}_0+\sum_q T^{(q)}_{0 }, \quad (q = \tfrac12, 1, \tfrac32,\dots)\label{eq:T_0}
\ee
where $T^{(t)}_0$ and $T^{(S)}_0$ are defined as
\be
T^{(t)}_0 =  \sum_{ij, \sigma} t^{\sigma\sigma}_{ij} f^\dagger_{i\sigma}  f_{j\sigma}, \quad T^{(S)}_0= \frac{J_K}{S}\sum_i a^\dagger_i a_i s^z_{i},\label{eq:T_0^tS}
\ee
%
%
and $ s^z_{i} = (f^\dagger_{i\up}f_{i\up}-f^\dagger_{i\down}f_{i\down})/2$ is the $z$ component of the electron spin.
The operator $T^{(t)}_0$ describes hopping of electrons with the same spin (in the local frame) and $T^{(S)}_0$ corresponds to an interaction of density-density type between the bosons and electrons. The latter comes from the Kondo coupling after substituting $S^z = S - a^\dagger_i a_i$. The operators which are labeled $T^{(q)}_{0 }$ all come from the usual $1/S$ expansion of the Heisenberg Hamiltonian $H_J$, see Eq.~\eqref{eq:H_J}, and thus only depend on the spin wave bosons. Each term $T^{(q)}_{0 }$ collects all contributions of order $S^{-q}$, and since $1/S$ expansions of Heisenberg models are standard, the precise form of $T^{(q)}_{0 }$ is not further discussed here but presented in Appendix~\ref{app:HP-J}. It is clear from the definition of $T_0$ that it indeed collects all terms which preserve the electron spin, i.e., all terms which are not spin-flip terms. (This is, of course, trivial for the contributions from $H_J$, since these do not depend on the fermions.)

Next, consider the operator $T_1 $. The operator $T_1$ collects all spin-flip terms which are of the type $\sim f^\dagger_{\up}  f_{\down}$ and therefore increase the angular momentum of the electrons by $1$. Specifically, $T_1$ is defined as
\be
T_1 = T^{(t)}_{1 } + \sum_q T^{(q)}_{1 }, \quad (q = \tfrac12, \tfrac32,\tfrac52,\dots)\label{eq:T_1}  ,
\ee
where $T^{(t)}_{1 }$ denotes the hopping term, which is given by
\be
T^{(t)}_{1 }  =  \sum_{ij} t^{\up\down}_{ij}  f^\dagger_{i\up}  f_{j\down} ,  \label{eq:T^t_1}
\ee
and the operators $T^{(q)}_{1 }$ denote the $S^{-q}$ terms coming from the $1/S$ expansion of $(J_K/2S) \sum_i S^-_i f^\dagger_{i\up}f_{i\down}$. For instance, the first two terms in this expansion, which correspond to $q=1/2$ and $q=3/2$, are given by
\beq
T^{(\frac12)}_{1 } &=&- \frac{J_K}{(2S)^{1/2}}\sum_i a^\dagger_i f^\dagger_{i\up}f_{i\down} , \label{eq:T^1/2_1}\\
 T^{(\frac32)}_{1 } &=& \frac{J_K}{2(2S)^{3/2}}\sum_i a^\dagger_i n_if^\dagger_{i\up}f_{i\down},\label{eq:T^3/2_1}
\eeq
Two types of spin-flip processes can thus be distinguished: ({\it i}) spin flips generated by hopping of electrons and ({\it ii}) spin flips generated by the interaction between the electrons and spin wave fluctuations of the local moments. In the case of the latter, fermion spin flips involve the emission of a spin wave. Spin flips originating from the hopping of electrons between sites $i$ and $j$ can only occur when the local moments at $i$ and $j$ are not perfectly aligned, such that $t^{\up\down}_{ij} \sim \bz^\dagger_i \bw_j$ is nonzero [see Eqs.~\eqref{eq:Omega} and \eqref{eq:t_ij_ss}].  

Since $T_{-1}$ is simply the Hermitian conjugate of $T_1$, the form of all $T_m$ is fully determined. It is straightforward to compute the commutators of $T_m$ with $H_0$, resulting in 
\be
[T_m,H_0] = mJ_K T_m. \label{eq:commutator}
\ee
This is intuitively sensible, since $T_{\pm1}$ are spin-flip operators and the energy difference between the spin-up and -down states is indeed $\pm J_K$. Instead, $T_0$ preserves spin and therefore commutes with $H_0$. 

More formally, the commutator \eqref{eq:commutator} shows that $T_m$ are eigenoperators of the Liouville operator $\mathcal L_0$, defined as $\mathcal L_0 A = [A,H_0]$ for any operator $A$ \cite{Becker:2002p235115,Sykora:2020p165122}. The operators $T_m$ are eigenoperators of $\mathcal L_0$ in the sense that
\be
\mathcal L_0 T_m = m J_K T_m, \label{eq:liouville}
\ee
where $m J_K$ are the eigenvalues. Since $H_0$ essentially measures the difference in the number of spin-up and spin-down electrons, the operators $T_{\pm 1}$ collect all terms which flip the electron spin (i.e., spin-flip terms), whereas $T_0$ collects all terms which preserve fermion flavor. 

\subsection{Canonical transformation \label{ssec:derivation}}

We are now in a position to define and derive the canonical strong coupling expansion. As mentioned previously, a key assumption is that $J_K$ is the largest energy scale, such that $ t /J_K$ and $J/J_K$ are small. The form of $H$ given by \eqref{eq:H_Tm} indeed reflects this, with $H_0$ simply describing a separation of the high- and low-energy degrees of freedom. We then seek a unitary transformation $e^{iQ}$, with Hermitian generator $Q$, such that the transformed Hamiltonian given by 
\be
\begin{split}
\mathcal H&=  e^{iQ} H e^{-iQ}, \\
 &= H + [iQ , H] + \frac{1}{2} [iQ,[iQ , H]] + \cdots, \label{eq:SW-transform}
\end{split} 
\ee
does not contain spin-flip processes (i.e., processes which change the number of occupied low- and high-energy states) up to a desired order in the small parameters $1/S$,  $ t /J_K$, and $J/J_K$. Below we describe an iterative prescription for determining $Q$.

This strong coupling expansion is inspired by other well-established perturbative approaches to quantum many-body systems based on some form of canonical Schrieffer-Wolff transformation~\cite{Schrieffer:1966p491,Bravyi:2011p2793,MacDonald:1988p9753,Nagaev:1998p827,Becker:2002p235115,Sykora:2020p165122}. In particular, the recipe for the determining $Q$ is formally equivalent to the $t/U$ expansion of the Hubbard model proposed in Ref.~\onlinecite{MacDonald:1988p9753}. The transformation obtained in Ref.~\onlinecite{MacDonald:1988p9753} eliminates terms that change the number of doubly occupied sites to any desired order in $t/U$, and thus yields a perturbative strong coupling expansion of the Hubbard model. In a similar fashion, here we aim to exploit the energetic separation of the two electron spin species, described by $H_0$, and remove all spin-flip terms. We note that more generally, the canonical spin wave expansion developed here fits into the framework of many-body diagonalization and renormalization schemes based on unitary transformations~\cite{Becker:2002p235115,Sykora:2020p165122}. 

The next step is to write $Q$ as a series expansion,
\be
Q= Q^{(1)}+Q^{(2)}+Q^{(3)}+\cdots, \label{Q-expand}
\ee
which should be understood as an expansion in the ``perturbation'', here given by $T_m$. This can be made explicit by formally replacing the terms $T_m$ with $\lambda T_m$, where $\lambda$ is a real parameter which functions as a bookkeeping device (much like in textbook treatments of perturbation theory). The terms $Q^{(p)}$ in \eqref{Q-expand} then collect all contributions proportional to $\lambda^p$. Since it is clear that the order $p$ simply corresponds to the number of times any of the operators $T_m$ appears, in what follows we will set $\lambda\rightarrow 1$. 
 
It is important to note at this point that the order $p$ does not---at least not \emph{directly}---correspond to the order of the small parameters in which ultimately seek an expansion. This follows from the fact that the operators $T_m$, as defined in Eqs.~\eqref{eq:T_0} and \eqref{eq:T_1}, are themselves expansions in $1/S$ and thus sums of terms of different order in $1/S$. The canonical spin wave expansion is therefore a nested or double expansion scheme (i.e., $1/S$ followed by strong coupling expansion), and some care must be taken when collecting terms of the same order in the effective Hamiltonian. In Sec.~\ref{ssec:H-expand} below we return to this point and describe how to account for the nested nature of the expansion scheme.

To proceed, we substitute \eqref{Q-expand} and \eqref{eq:H_Tm} into \eqref{eq:SW-transform} and obtain 
 \begin{multline}
 \mathcal H= H_0 +T_0 +T_1+T_{-1} + \left[iQ^{(1)} , H_0\right] + \left[iQ^{(1)} , T_0\right]+ \\ 
+  \left[iQ^{(1)} , H_0\right]+ \left[iQ^{(1)} , T_1+T_{-1}\right]  +\left[iQ^{(2)} ,H_0\right] \\
+ \frac{1}{2} \left[iQ^{(1)},\left[iQ^{(1)} , H_0\right]\right] +  \cdots, \label{eq:H-expand}
 \end{multline}
from which $Q^{(p)}$ can be determined successively by requiring that all spin-flip terms of order $p$ on the right hand side are removed~\cite{MacDonald:1988p9753}. For instance, at lowest order, $Q^{(1)}$ is determined by the condition
\be
 [iQ^{(1)}, H_0] +T_1 + T_{-1} = 0 , \label{eq:Q1-def}
\ee
from which one finds
\be
Q^{(1)} = i(T_1 - T_{-1} )/J_K, \label{eq:Q1}
\ee
by virtue of the commutator \eqref{eq:commutator}. At general order $p$, all terms on the right hand side of Eq.~\eqref{eq:H-expand} are of the form
\be
T^{(p)}_{\bf m} \equiv T_{m_1}T_{m_2}\ldots T_{m_p}, \label{T^p_m}
\ee
and those terms for which $\sum_i m_i \neq 0$ can be removed by an appropriate choice of $Q^{(p)}$. The transformed Hamiltonian $\mathcal  H$ then consists of all terms $T^{(p)}_{\bf m}$ for which $\sum_i m_i = 0$ and thus do not represent spin flip terms. The transformed Hamiltonian $\mathcal H$ may be formally expanded as
\be
{\mathcal H } = {\mathcal H }^{(0)}+ {\mathcal H }^{(1)} +{\mathcal H }^{(2)} +{\mathcal H }^{(3)} + \cdots , \label{eq:H-expansion} 
\ee
such that $ {\mathcal H}^{(p)}$ collects all contributions at order $p$. At lowest order $p=0,1$, we simply have ${\mathcal H}^{(0)} = H_0$ and ${\mathcal H}^{(1)} = T_0 $, and for the next two orders we find
\beq
{\mathcal H}^{(2)} &=& -[T_1,T_{-1}]/J_K \label{eq:H(2)} \\
{\mathcal H}^{(3)} &=& \left([ [T_1,T_{0}],T_{-1}] +\text{H.c.}\right)/2J^2_K, \label{eq:H(3)} 
\eeq
in agreement with Ref.~\onlinecite{MacDonald:1988p9753}. Since the degree of commutator nesting increases at each order, the contributions to ${\mathcal H }$ become more involved as the order increases. The expansion can be carried out straightforwardly to any desired order, although evaluation of the commutators may be become laborious. For the purpose of this work, we have determined the transformed Hamiltonian up to and including ${\mathcal H}^{(5)}$. In Appendix~\ref{app:canonical} we collect some more details on the derivation of ${\mathcal H}^{(3)}$ in Eq.~\eqref{eq:H(3)}, as well the determination of ${\mathcal Q}^{(2)}$ and ${\mathcal Q}^{(3)}$.

\subsection{Determination of effective Hamiltonian \label{ssec:H-expand}}

By construction, the canonical transformation has removed the spin-flip processes and as a result, the transformed Hamiltonian of Eq.~\eqref{eq:H-expansion} no longer connects sectors with different electron spin quantum numbers. By projecting out the high-energy electrons one then obtains an effectively spinless Hamiltonian for the low-energy electrons and the spin wave bosons, which is an expansion in the small parameters $1/S$, $t/J_K$, and $J/J_K$. To collect terms of the same order in the expansion parameters, one has to account for the fact that the strong coupling expansion essentially constitutes a double expansion: a $1/S$ expansion of the spins, followed by a strong coupling expansion in $1/J_K$. Therefore, appropriately assembling the effective Hamiltonian is achieved by substituting the form of $T_m$ as given by \eqref{eq:T_0} and \eqref{eq:T_1} into Eq.~\eqref{eq:H-expansion} and then collecting terms of the same order. In the final part of this section we briefly describe this procedure. 

Consider the second order contribution to the transformed Hamiltonian given by Eq.~\eqref{eq:H(2)}. The operators $T_{\pm 1}$ are given by Eq.~\eqref{eq:T_1}, but in practice only the lowest order terms in the expansion of $T_{\pm 1}$ are retained, depending on the desired expansion in $1/S$. For instance, if one wishes to include quantum corrections to linear spin wave order, i.e., retain terms of order $1/S^2$, it is necessary to take 
\be
T_{\pm 1}\simeq   T^{(t)}_{\pm 1 } + T^{(\frac12)}_{\pm1 }+ T^{(\frac32)}_{\pm1 } \label{eq:T_1-approx}
\ee
For the moment, however, let us omit the last $q=3/2$ term and focus on the first two terms. The commutator $[T_1,T_{-1}]$ then takes the form
\begin{multline}
[T_1,T_{-1}] = \left[T^{(t)}_1,T^{(t)}_{-1}\right]+ \left[T^{(\frac12)}_1,T^{(\frac12)}_{-1}\right]+ \left[T^{(t)}_1,T^{(\frac12)}_{-1}\right] \\
+\left[T^{(\frac12)}_1,T^{(t)}_{-1}\right]. \label{com-expand}
\end{multline}
Let us examine what each term on the right hand side represents. First, since $T^{(t)} \sim t$, the first term gives a contribution to the Hamiltonian of order $t^2/J_K$, and hence represents a $t/J_K$ correction to the classical strong coupling limit ($J_K \rightarrow \infty$) of the Kondo Hamiltonian. Similarly, since $T^{(\frac12)}_{\pm 1} \sim S^{-1/2}$, the second term gives a contribution of order $J_K/S$, and thus corresponds to a $1/S$ correction of the Kondo coupling. (We discuss the significance of this term in detail in the next section.) The remaining two terms on the right hand side of \eqref{com-expand} are of order $t/S^{-1/2}$ and will contribute to linear spin wave theory in the second order diagrammatic calculation of the boson self-energy. 

This example of the commutator $[T_1,T_{-1}]$ clarifies how ${\mathcal H}^{(2)}$ contains contributions of different type and shows how this is due to the structure of the $T_m$ operators. It also shows that contributions of the same order in, for instance, $1/S$ must be collected correctly. For instance, had we retained the $q=3/2$ term in \eqref{eq:T_1-approx}, then \eqref{com-expand} would have produced terms of order $t/S^{-3/2}$, as is easily seen. Terms of this order are also contained in ${\mathcal H}^{(3)}$ and must therefore be collected properly. In Appendix \ref{app:H(3)} we consider ${\mathcal H}^{(3)}$, given by \eqref{eq:H(3)}, in detail and show how relevant contributions to the effective Hamiltonian are appropriately collected. 

After properly accounting for the nested nature of the expansion, the result of the canonical transformation is an effective Hamiltonian for the low-energy dynamics of itinerant magnets in the strong coupling regime. In Sec.~\ref{sec:H_eff} we present and discuss the form of this Hamiltonian in detail. 


\section{Discussion of the canonical spin wave expansion \label{sec:discussion}}

Before presenting the effective Hamiltonian in the next section, in this section we aim to provide deeper insight into the nature and interpretation of the canonical transformation derived in the previous section. We will address three aspects in particular. First, we will show that the canonical transformation can be viewed as an order-by-order diagonalization of the quantum Kondo atom considered in Sec.~\ref{sec:klm} [See Eqs.~\eqref{eq:H_Ki} and \eqref{eq:eps_pm}]. We will then discuss the nature of the transformed boson and fermion operators, which we will see is closely related to the fact that the canonical transformation diagonalizes the Kondo coupling. A third and final important aspect is the connection between the canonical spin wave expansion presented here and previous approaches to spin wave expansions of Kondo lattice magnets.

\subsection{Diagonalization of Kondo coupling \label{ssec:kondo-diag}}

A first important insight into the structure of the canonical transformation can be gained by considering the special case $t_{ij} = J_{ij} = 0$, i.e., keeping only the local Kondo coupling. Computing the transformed Hamiltonian with the appropriately simplified form of $T_m$ yields
\beq
\mathcal H  &=  & H_0 + \mathcal H^{(1)} +\mathcal H^{(2)} +\ldots  \label{eq:H_K-only} \\
&\simeq & H_0 + T^{(S)}_0 - \frac{1}{J_K}\left[T^{(\frac{1}{2})}_1,T^{(\frac{1}{2})}_{-1}\right]+\ldots, \label{eq:H_K-only-expand} 
\eeq
where we have listed only the lowest order terms. Evaluating the commutator, we find that $\mathcal H$ takes the form
\be
\mathcal H =  - J_K \sum_i \tilde s^z_{i}  +\frac{J_K}{2S} \sum_{i}\tilde n_{i\down} (1  -\tilde n_{i\up}), \label{eq:H_K_transform}
\ee
where $\tilde n_{i\sigma} = \tilde f^\dagger_{i\sigma} \tilde f_{i\sigma} $. Here we use a tilde to denote the transformed electron operators, i.e., the electron operators after canonical transformation (see Sec.~\ref{ssec:op-expand}). (Distinguishing these operators notationally will serve the discussion of the interpretation of the transformed operators.) The first term in \eqref{eq:H_K_transform} is just the classical Kondo coupling $H_0$, and the second term can be viewed as the $1/S$ correction to the classical Kondo coupling. 

The key observation is that the transformed Hamiltonian \eqref{eq:H_K_transform} exactly reproduces the correct \emph{quantum} eigenvalues of the local Kondo coupling. To see this, consider the states 
\be
\ket{2_i} = \tilde f^\dagger_{i\up} \tilde f^\dagger_{i\down}\ket{0},\; \ket{\up_i}=  \tilde f^\dagger_{i\up} \ket{0},\; \ket{\down_i}= \tilde f^\dagger_{i\down} \ket{0},
\ee
on a single site $i$, which together with the vacuum $\ket{0}$ span the local Hilbert space. These four states are eigenstates of \eqref{eq:H_K_transform}. The two states $\ket{0}$ and $\ket{2_i} $ have zero energy, whereas $\ket{\up_i}$ and $\ket{\down_i}$ have energies $-J_K/2$ and $J_K(S+1)/2S$, respectively. These are precisely the energies given in Eq.~\eqref{eq:eps_pm} and would seem to imply that the transformed Hamiltonian given by \eqref{eq:H_K_transform} is an exact representation of the quantum Kondo coupling. 

A natural question, however, concerns the fate of the higher order corrections generated by the canonical transformation but not listed in \eqref{eq:H_K-only-expand}. Remarkably, we find that these higher order corrections vanish at all orders. For instance, at order $1/S^2$ we find that
\begin{multline}
- \frac{1}{J_K}\left( \left[T^{(\frac32)}_1,T^{(\frac12)}_{-1}\right]+\left[T^{(\frac12)}_1,T^{(\frac32)}_{-1}\right] \right) + \\
+ \frac{1}{2J^2_K}\left( \left[\left[T^{(\frac12)}_1, T^{(S)}_0\right],T^{(\frac12)}_{-1}\right]+\text{H.c.} \right) \\
+ \frac{1}{4J^3_K}\left( \left[\left[\left[T^{(\frac12)}_1,T^{(\frac12)}_{-1}\right], T^{(\frac12)}_{1}\right],T^{(\frac12)}_{-1}\right]+\text{H.c.} \right) \\
=0,
\end{multline}
showing that \eqref{eq:H_K_transform} is indeed  the transformed Kondo coupling up to and including order $1/S^2$. That higher order corrections vanish might have been anticipated given that \eqref{eq:H_K_transform} already reproduces the exact quantum eigenvalues. Our explicit check confirms this.

We thus arrive at the conclusion that when $t_{ij} = J_{ij} = 0$, the transformed Hamiltonian $\mathcal H$ is indeed an exact representation of the quantum Kondo atom. Another way of stating this it to note that the canonical transformation may be viewed as an order-by-order diagonalization scheme for the Kondo coupling. This is perhaps not as surprising, given that the canonical transformation is a strong coupling expansion, i.e., an expansion in the inverse Kondo coupling $J_K$, which is assumed to be the largest energy scale. In this sense it is natural to expect that the canonical transformation would amount to some form of diagonalization of the principal term in the Hamiltonian.

\subsection{Transformed boson and fermion operators \label{ssec:op-expand}}

Next, consider the transformed fermion and boson operators. These operators are given by the inverse transformation $\tilde f_{i\sigma}  = e^{-iQ}  f_{i\sigma}  e^{iQ} $ and $\tilde a_{i}  = e^{-iQ}  f_{i}  e^{iQ} $, and can be expanded in terms of the bare untransformed operators once $Q$ is determined using the recipe described in Secs.~\ref{ssec:derivation} and \ref{ssec:H-expand}. We will first discuss the transformed electron operators and then the boson operators.

For the electron operators we have just seen that these should be interpreted as creating or annihilating electrons in a state of total spin $S\pm 1/2$. Listing only the lowest order terms, and ignoring any contributions from the Heisenberg coupling for the moment, we find
\begin{widetext}
\begin{align}
\tilde f_{i\up} &= f_{i\up} + \frac{a^\dagger_i f_{i\down}}{\sqrt{2S}}  - \frac{f_{i\up}}{4S} (f^\dagger_{i\down}f_{i\down} + a^\dagger_i a_i )-  \frac{1}{J_K}\sum_{j} \left[ t^{\up\down}_{ij}f_{j\down}+  \frac{1}{\sqrt{8S}}\left( t^{\down\up}_{ij}  a^\dagger_{i}f_{j\up}+ t^{\up\down}_{ij}a_{j}f_{j\up} -2t^{\up\up}_{ij}  a^\dagger_{j}f_{j\down} +2t^{\down\down}_{ij}a^\dagger_{i}f_{j\down} \right)\right] + \cdots, \label{eq:f_up-transform} \\
\tilde f_{i\down} &= f_{i\down} - \frac{a_i f_{i\up} }{\sqrt{2S}} - \frac{f_{i\down}}{4S} (f_{i\up}f^\dagger_{i\up} + a^\dagger_i a_i ) + \frac{1}{J_K}\sum_{j}\left[ t^{\down\up}_{ij}f_{j\up} -\frac{1}{\sqrt{8S}} \left( t^{\up\down}_{ij}  a_{i}f_{j\down}+ t^{\down\up}_{ij}a^\dagger_{j}f_{j\down} +2t^{\up\up}_{ij}  a_{i}f_{j\up} -2t^{\down\down}_{ij}a_{j}f_{j\up} \right)\right]+ \cdots. \label{eq:f_down-transform}
\end{align}
\end{widetext}
Here we have kept terms up to order $1/S$ and have included the lowest order $t/J_K$ correction to the $S^{-1/2}$ terms. Up to these orders in the expansion parameters Eqs.~\eqref{eq:f_up-transform} and \eqref{eq:f_down-transform} express the transformed fermion operators in terms of the original electronic degrees of freedom (in the local frames). To see how this corroborates the interpretation of the transformed fermion operators, consider a single site $i$ and temporarily set $t_{ij}=0$. Then let $\tilde f^\dagger_{i\sigma}$ act on the state $\ket{0}_i \equiv \ket{S,S}_i$, i.e., the maximal weight state with no bosons. For the up-fermions ($\sigma=\up$) we simply find $\tilde f^\dagger_{i\up}\ket{0}_i = f^\dagger_{i\up}\ket{0}_i $, whereas for the down-fermions ($\sigma=\down$) we obtain
\be
\tilde f^\dagger_{i\down} \ket{0}_i= \left(1 - \frac{1}{4S}\right) f^\dagger_{i\down}\ket{0}_i - \frac{1}{\sqrt{2S}} a^\dagger_i f^\dagger_{i\up} \ket{0}_i+\cdots . \label{eq:f_down,i}
\ee
These expressions for the states $\tilde f^\dagger_{i\sigma}\ket{0}_i $ should then be compared to the states $\ket{S\pm\tfrac12,S\pm\tfrac12}$ given in Eqs.~\eqref{eq:psi_+} and \eqref{eq:psi_-}. (Here we have set $M= S+1/2$ and $M=S-1/2$, respectively.) Recall that the latter are states of total spin $S\pm1/2$ and are eigenstates of the quantum Kondo coupling. Expanding the coefficients $u$ and $v$ of Eq.~\eqref{eq:u,v} in $1/S$, and recognizing that $a^\dagger_i  \ket{0}_i = \ket{S,S-1}_i$, shows that the states $\tilde f^\dagger_{i\sigma}\ket{0}_i $ are indeed properly understood as states of total spin $S\pm 1/2$. Hence, a detailed exam of the transformed electron operators confirms what the discussion of Sec.~\ref{ssec:kondo-diag} seemed to indicate.

Let us next consider the first order $t/J_K$ contributions in Eqs.~\eqref{eq:f_up-transform} and \eqref{eq:f_down-transform}, which reflect the strong coupling expansion produced by the canonical transformation. Two observations can be made. We first notice that these contributions cause the transformed fermion operators to depend on the ground state configuration of the local moments. This follows from the structure of the hoppings $t^{\sigma\sigma'}_{ij}$, which depend on the angles $\{ \theta_i, \phi_i\}$ and thus encode the classical ground state. The second observation is that the $t/J_K$ contributions make the transformed fermion operators non-local in terms of the original degrees of freedom. This is natural and expected, since the hoppings $t_{ij}$ connect neighboring sites. 

Two types of $t/J_K$ contributions can be distinguished in Eqs.~\eqref{eq:f_up-transform} and \eqref{eq:f_down-transform}. The first type is of zeroth order in $1/S$ and hence is already present at the classical level. Terms of this type simply arise from the perturbative removal of hoppings between the up and down spin states, and are therefore proportional to the effective spin-off-diagonal hoppings $t^{\sigma\bar\sigma}_{ij}$. Terms of the second type are of order $S^{-1/2}$ and therefore reflect the quantum nature of the spins. Indeed, these terms feature fermions dressed with bosons, which describe spin flips, and can be understood as $t/J_K$ corrections to the local $S^{-1/2}$ term, i.e., second term in Eqs.~\eqref{eq:f_up-transform} and \eqref{eq:f_down-transform}. The appearance of electrons ``dressed'' with non-local spin flip excitations suggests that the transformed fermion operators can be interpreted as a form of spin polarons. 

\begin{figure}
	\includegraphics[width=0.8\columnwidth]{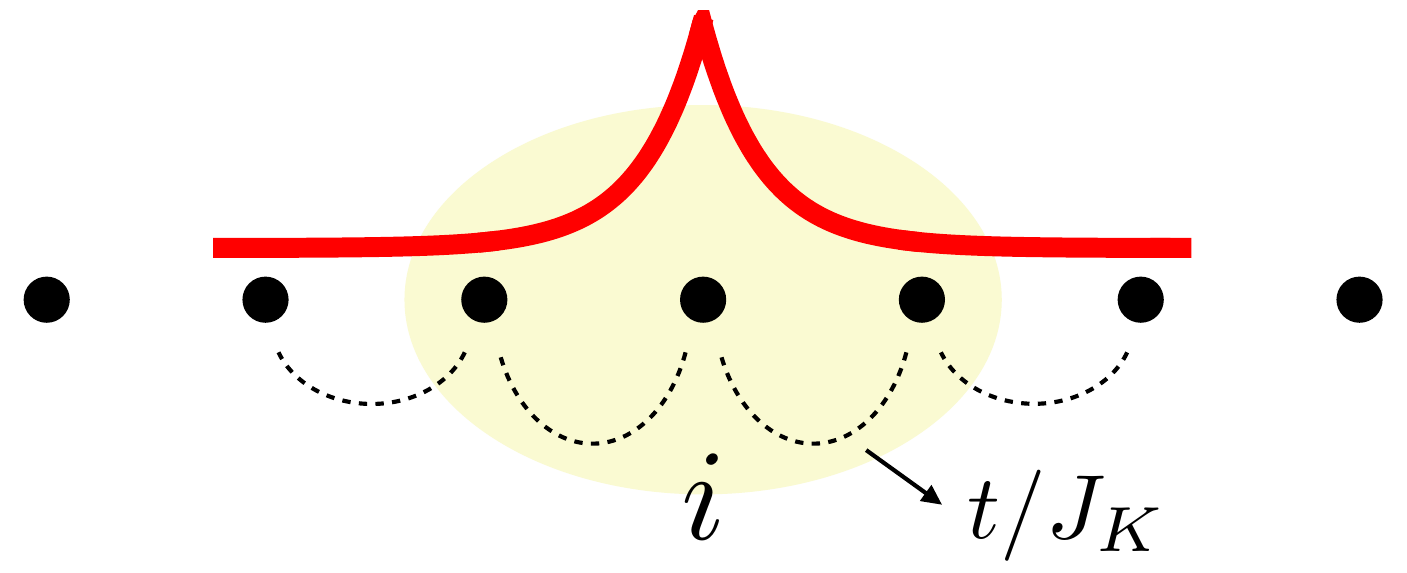}
	\caption{Schematic representation of the electron operators in Eq.~\eqref{eq:f-down-polaron}, which create spin polarons centered at site $i$ with finite spatial extent. The red curve indicates the spatial spread of the spin polaron states.}
	\label{fig:polaron}
\end{figure}

To examine this connection with spin polaron physics in more detail, and to demonstrate that this is indeed a valid interpretation, it is instructive to consider the form of the transformed fermion operators in the specific case of a ferromagnet. In the case of a ferromagnet, the effective hoppings simplify and one has $t^{\sigma\sigma'}_{ij} = t_{ij}\delta_{\sigma\sigma'}$. It is in particular insightful to consider the transformed down-fermions, which take the form
\be
\tilde f^\dagger_{i\down} = f^\dagger_{i\down} + \frac{1}{\sqrt{2S}}\sum_{jk} \Phi_{ijk} a^\dagger_j f^\dagger_{k\up} + \cdots. \label{eq:f-down-polaron}
\ee
Here we have only retained the $S^{-1/2}$ terms and have defined $ \Phi_{ijk} $ as 
\begin{multline}
 \Phi_{ijk}  = \delta_{ij}\delta_{ik} + \frac{1}{J_K} (t_{ij}\delta_{jk}-\delta_{ij}t_{jk})  \\
-  \frac{1}{J^2_K}[ 2t_{ij}t_{jk} -  \sum_l (t_{il}t_{lj}\delta_{jk}+t_{il}t_{lk}\delta_{ij})] . \label{eq:Phi_ijk}
\end{multline}
Note that, for emphasis, here we have included the order $t^2/J^2_K$ corrections, which are not included in Eq.~\eqref{eq:f_down-transform}. The transformed operators of Eq.~\eqref{eq:f-down-polaron} create electrons which, in terms of the original degrees of freedom, clearly have the structure of a down-electron dressed with spin-flip excitations. As is apparent from \eqref{eq:Phi_ijk}, when the $t/J_K$ corrections are ignored, Eq.~\eqref{eq:f-down-polaron} simply reduces to the two lowest order terms of \eqref{eq:f_down,i}. This limit thus corresponds to a purely ``on-site'' spin polaron with no spatial extent, which is nothing but an electron in a spin $S-1/2$ state with one local moment. Including the $t/J_K$ corrections introduces non-local terms and gives rise to finite extent of the spin polaron. In particular, the extent of the cloud of spin-flip excitation increases as higher order $t/J_K$ corrections are taken into account. We therefore find that, in the specific case of the ferromagnet, the operators \eqref{eq:f-down-polaron} can indeed be viewed as creating electrons in a state of total spin $S-1/2$ centered at site $i$, but these states have finite extent when expressed in terms of the bare untransformed degrees of freedom. The form of \eqref{eq:Phi_ijk} implies that the polaron wave function decays as $\sim (t/J_K)^{i-j}$, which suggests that in the strong coupling regime the size of the polaron $\xi$ behaves as $\xi^{-1} \sim \ln (t/J_K)$. This is schematically depicted in Fig.~\ref{fig:polaron}. Only in the strong coupling limit, when $J_K \rightarrow \infty$, do these states become strictly local on-site objects of total spin $S-1/2$. 

It is worth stressing that the spin polaron structure is a quantum phenomenon and disappears in the classical limit defined by $S \rightarrow \infty$.  

Our finding that the transformed fermion operators should be viewed as spin polarons is in perfect agreement with previous studies of Kondo lattice ferromagnets, in particular exactly solvable toy models in one dimension~\cite{Ueda:1992p1030,Tsunetsugu:1997p809}. This connection with previous work on spin polaron physics will be explored in more detail elsewhere. 


Finally, we turn to the transformed boson operators given by $\tilde a_{i}  = e^{-iQ}  a_{i}  e^{iQ} $. Again, excluding contributions from the Heisenberg couplings $H_J$ we find that the transformed boson operators take the form
\begin{multline}
\tilde a_i =  a_i + \frac{1}{\sqrt{2S}} s^+_i- \frac{1}{2S}a_i s^z_i  +  \frac{1}{J_K\sqrt{8S}}\sum_{j}\left( t^{\down\up}_{ji}  f^\dagger_{j\down}f_{i\down} \right. \\
\left. -t^{\down\up}_{ij}f^\dagger_{i\up}f_{j\up} +2t^{\up\up}_{ji}  f^\dagger_{j\up}f_{i\down} -2t^{\down\down}_{ij}f^\dagger_{i\up}f_{j\down} \right) +\cdots, \label{eq:a-transform}
\end{multline}
where, as before, we have included terms up to order $1/S$ and included $t/J_K$ corrections to the $S^{-1/2}$ terms.  The form of the transformed boson operators in \eqref{eq:a-transform} suggests that these should be interpreted as spin-flip operators of the total spin at site $i$, i.e., the quantum mechanical sum of the of local moment and electron spin. This follows from the presence of the $s^+_i$ term in the right hand side, which corresponds to an electron spin flip. The terms which are of order $t/J_K$ indicate that the electron spin flip need not be local. 

To examine this interpretation of the boson operators further, we consider the form raising operator of total spin at site $i$, which is given by $T^+_i = S^+_i + f^\dagger_{i\up}f_{i\down} $, when expressed in terms of the transformed operators $\tilde a_i $ and $\tilde f_{i\sigma} $. Expanding $S^+_i $ in the bosons $a_i$, and then transforming all operators according to $a_i \rightarrow e^{iQ} a_i e^{-iQ}$ (similarly for the electrons), we find that $T^+_i$ takes the form
\begin{multline}
S^+_i + f^\dagger_{i\up}f_{i\down}  =\sqrt{2S}  \tilde a_i \bigg[ 1+ \frac{ \tilde s^z_i}{2S}   \\
+ \frac{1}{4J_KS}\sum_j t_{ij} \big(\tilde f^\dagger_{i\up}\tilde f_{j\up} - \tilde f^\dagger_{i\down}\tilde f_{j\down} - \{ i\leftrightarrow j\}\big) \bigg] \\
+\frac{1}{J_K}\sum_j t_{ij} (\tilde f^\dagger_{i\up}\tilde f_{j\down}-\tilde f^\dagger_{j\up}\tilde f_{i\down}) + \cdots.
\end{multline}
We see from this form that $ \tilde a_i$ indeed should be interpreted as the raising operator for total spin, and hence corresponds to fluctuation of the total spin at site $i$. The multiplicative factor which depends on the fermion operators reflects the fact that total spin at site $i$ is an operator quantity and depends on the presence of electrons. The total spin can takes values $S\pm 1/2$, but also $S$ when no or two electrons occupy site $i$, and may be represented as $T = S  +  \tilde s^z_i$. 

The notion that the transformed boson operators $ \tilde a_i$ describe fluctuations of an operator-valued total spin $T =S  +  \tilde s^z_i$ is very much consistent with earlier work by Shannon and Chubukov~\cite{Shannon:2002p104418}, a connection we discuss in more detail below in Sec.~\ref{ssec:previous-work}.

To conclude this discussion of the transformed operators we summarize the main takeaways. We have found that the strong coupling spin wave expansion of the Kondo lattice model, achieved via canonical transformation, is expressed in terms of natural degrees of freedom of the strong coupling regime. These correspond to electrons in a state of total spin $S\pm 1/2$ with the local moments (i.e., electron states in the eigenbasis of the Kondo coupling) and spin wave bosons which describe spin flips of the operator-valued total spin on a given site. The implication is that this strong coupling spin wave expansion correctly captures the quantum properties of the local moment spins order-by-order in $1/S$.

\subsection{Relation to previous work \label{ssec:previous-work}}

Following the preceding discussion of the nature of the canonical spin wave expansion, we conclude this section by discussing its connection to previous work on spin wave expansions in Kondo lattice magnets. 

As mentioned in the introduction, most studies of spin wave dynamics in Kondo lattice magnets have focused on ferromagnets~\cite{Kubo:1972p21,Furukawa:1996p1174,Nagaev:1998p827,Perkins:1999p1182,Golosev:2000p3974,Shannon:2002p104418}. To obtain the magnon dispersion of Kondo ferromagnets to the leading order in $1/S$, it is sufficient and straightforward to simply perform a standard spin wave expansion of the local moment spins and compute the appropriate boson self-energy diagrams~\cite{Kubo:1972p21,Furukawa:1996p1174}. In the context of the expansion developed here, this amounts to computing the relevant diagrams based on the bare \emph{untransformed} Hamiltonian given by Eq.~\ref{eq:H_local}. (In the case of ferromagnets a rotation to a local frame is not necessary.) Including quantum corrections beyond leading order is significantly more challenging, however, as the diagrammatics becomes impractical if not unfeasible. This point was emphasized in Ref.~\onlinecite{Shannon:2001p6371}.

One approach to addressing this challenge is to remove the terms linear in the bosons by means of a canonical transformation. This was proposed in Ref.~\onlinecite{Nagaev:1998p827}. As recognized in Ref.~\onlinecite{Nagaev:1998p827}, the presence of terms linear in the bosons signals that the bosons do not describe the true magnons, since one expects the number of (true) magnons to be conserved in ferromagnets. The proposed canonical transformation was therefore interpreted as a way to express the Hamiltonian in terms of the true magnons, i.e., the fluctuations of the sum of electron and local moment spin. Since the resulting Hamiltonian better reflects the underlying physics, especially in the strong coupling regime, it overcomes some of the technical difficulties associated with computing quantum corrections to spin wave properties in leading order~\cite{Golosev:2000p3974}. Such corrections are of particular interest, since the leading order linear spin wave dispersion of a Kondo ferromagnet in the strong coupling limit ($J_K \rightarrow \infty$) is equivalent to that of a Heisenberg ferromagnet. Quantum corrections show significant deviations from a Heisenberg ferromagnet, however, and thus expose the fundamental differences between Kondo and Heisenberg ferromagnets~\cite{Golosev:2000p3974,Wurth:1998p403,Okabe:1997p21,Shannon:2002p104418}.

The transformation first proposed in Ref.~\onlinecite{Nagaev:1998p827} is similar in spirit to the canonical transformation derived in this work, but quite different in its conception and structure. First, it is specifically designed to apply to a ferromagnet, although generalizations to other ordered states such as antiferromagnets are possible~\cite{Lv:2010p045125}. Second, and more importantly, the canonical transformation of Ref.~\onlinecite{Nagaev:1998p827} is not constructed iteratively by demanding that couplings between up and down fermions are removed at each order in the expansion, but is instead constructed by the mere requirement that the terms linear in the bosons are removed. The benefit is that such construction does not rely on the assumption of strong coupling. The downside is that the resulting Hamiltonian does not constitute a systematic and physically transparent $1/S$ expansion of the Kondo lattice Hamiltonian. A consequence of the latter is that the transformed boson operators do not, in fact, correspond to the true magnons in the limit of strong coupling. To see this, it is instructive to consider the unitary transformation of Ref.~\onlinecite{Nagaev:1998p827} in the strong coupling limit, and compare it to our unitary transformation $e^{iQ}$ as defined by Eqs.~\eqref{eq:SW-transform} and \eqref{Q-expand}. This comparison shows that the former coincides with the latter only when $Q$ is taken equal to $Q^{(1)} $ given by Eq.~\eqref{eq:Q1}, with $T_{\pm 1}$ given by \eqref{eq:T^1/2_1}. That is to say, only when $Q$ is restricted to the leading order contribution. This affects the form of the transformed Hamiltonian and the transformed fermion and boson operators beyond the leading order contributions, and in particular is the reason that the transformed magnons are only leading order approximations of the true magnons in the strong coupling limit. Our canonical transformation is therefore not the same has the transformation of Ref.~\onlinecite{Nagaev:1998p827}.


Our spin wave expansion is most closely related to the work of Shannon and Chubukov~\cite{Shannon:2002p104418,Shannon:2001p6371}, who proposed a systematic $1/S$ expansion of the Kondo lattice model in the double-exchange limit---here more generally referred to as the strong coupling limit. Their starting point was the intuition that in the double-exchange limit, it is more natural to formulate the problem in terms of the total spin of the electron and the local moment, since it is the eigenstates of total spin which diagonalize the on-site Kondo coupling. By introducing fermion and boson operators corresponding to the total spin on a given site, their ingenuity led them to derive a transformation between these operators and the constituent electron and local moment operators. As a result, their transformed boson operators describe fluctuations of the total spin on a given site and thus---by design---correspond to the true strong coupling magnons. We have similarly demonstrated that the transformed operators given by Eqs.~\eqref{eq:f_up-transform}, \eqref{eq:f_down-transform}, and \eqref{eq:a-transform}, should be interpreted as fermions creating or annihilating a total spin $S\pm 1/2$ state, and bosons creating or annihilating excitations of that total spin. In fact, we find that setting $t_{ij} = 0$ in Eqs.~\eqref{eq:f_up-transform}, \eqref{eq:f_down-transform}, and \eqref{eq:a-transform} exactly reproduces the transformed fermion and boson operators obtained by Shannon and Chubukov. 

This remarkable agreement shows that our strong coupling expansion can be viewed as a formalization and extension of the spin wave expansion proposed by Shannon and Chubukov. In particular, our strong coupling expansion of the Kondo lattice Hamiltonian derived in Sec.~\ref{ssec:derivation} produces not only a systematic spin wave expansion in $1/S$, but also a perturbative expansion in $t/J_K$ and $J/J_K$. This is compelling upshot of the strong coupling expansion via canonical transformation. It therefore offers a systematic way of including $t/J_K$ corrections to the strong coupling limit. This is clearly reflected in the structure of the transformed operators given by Eqs.~\eqref{eq:f_up-transform}, \eqref{eq:f_down-transform}, and \eqref{eq:a-transform}, which become non-local and thus account for the spin polaron physics originating from spin-flip processes. A detailed analysis of the way in which our strong coupling expansion capture the spin polaron nature of the fermions will be presented elsewhere. Our expansion furthermore treats the direct Heisenberg exchange coupling between the local moments on the same footing as the indirect exchange mediated by the charge carriers.

\section{Effective low-energy Hamiltonian \label{sec:H_eff}}

We now turn to the central result of the strong coupling spin wave expansion, the effective low-energy Hamiltonian. As emphasized in Sec.~\ref{sec:transform}, this Hamiltonian is effectively spinless after projection onto the low-energy electron states. An important observation is that the form of the effective Hamiltonian depends crucially on the sign of the Kondo coupling $J_K$. This follows from the discussion of Sec.~\ref{sec:discussion}, which has shown that the electron operators after canonical transformation correspond to electrons in a state of total spin $S\pm1/2 $ with the local moment. As detailed in Sec.~\ref{sec:klm}, the quantum mechanical structure of such electron states is rather different, and this is reflected in the form of the effective Hamiltonian. In particular, the effective Hamiltonian obtained for spin-up and spin-down electrons (which should be read as $S\pm1/2 $ electrons) is not simply related by changing $J_K \rightarrow -J_K$. For concreteness and simplicity, in this section we will primarily consider the case $J_K>0$, in which case the $S+1/2$ states are the low-energy states. The opposite case $J_K<0$ will be considered in the final part of this section (Sec.~\ref{ssec:f-down}), where the key differences between the two cases will be highlighted.

To examine the various terms in the effective Hamiltonian some remarks on notation and structure are necessary. The effective Hamiltonian can broadly be partitioned into three groups of terms, which we denote $\mathcal H_f$, $\mathcal H_{a}$, and $\mathcal H_{f,a}$. Here $\mathcal H_f$ and $\mathcal H_{a}$ collect all terms which only contain electron operators ($f$) and boson operators ($a$), and $\mathcal H_{f,a}$ collects terms which couple the electrons and bosons. Each of these three Hamiltonian terms constitute an expansion in $1/S$, $t/J_K$, and $J/J_K$. Taking $\mathcal H_f $ as example, we express this as
\be
\mathcal H_f = \sum_{mnl} \mathcal H_f^{(m,n,l)}, \label{eq:H_mnl}
\ee
where $ \mathcal H_f^{(m,n,l)}$ is of order $S^{-m}t^nJ^l/J^{n+l}_{K}$. In the remainder of this section we examine the lowest order terms for each of the three parts of the Hamiltonian. We will proceed in three stages. First, we focus on the classical limit given by $m=0$ and determine the $t/J_K$ corrections to the strong coupling limit ($J_K\rightarrow \infty$). Next, we consider the lowest order $1/S$ terms, given by $m=1$, of which we examine two types: quantum corrections to the fermion dynamics (i.e., terms part of $\mathcal H_f $) and linear spin wave terms (i.e., terms part of $\mathcal H_a $ and $\mathcal H_{f,a} $). We then proceed and consider quantum corrections to the linear spin wave terms, which are terms in $\mathcal H_a $ and $\mathcal H_{f,a} $ of order $1/S^2$, corresponding to $m=2$. Such terms encode the quantum nature of the local moments beyond linear spin wave order.

As mentioned, we conclude this section by discussing the key differences that arise when the Kondo coupling is assumed to be antiferromagnetic ($J_K<0$ in our convention).

A final remark about notation is in order. In the case of ferromagnetic Kondo coupling the low-energy electron operators are given by the transformed operators $\tilde f_{i\up}$, but in what follows we drop the spin label and the tilde, and simply write $f_i$ instead of $\tilde f_{i\up}$. All electron operators $f_i$ should be understood as transformed operators describing electrons in a state of total spin $S+1/2$ with the local moments.

\subsection{Classical strong coupling limit and $t/J_K$ corrections \label{ssec:classical}}

In the classical strong coupling limit the Hamiltonian for the electrons is quadratic and takes the form
\be
\mathcal H_f^{(0,0,0)}  =  \sum_{ij} t^{\up\up}_{ij} f^\dagger_i f_j, \label{eq:H_DE}
\ee
where $t^{\up\up}_{ij} = t_{ij}  \Omega^{\up\up}_{ij}$ describes the hopping of electrons in a background of ordered local moments, as defined in Eq.~\eqref{eq:t_ij_ss}. The $t/J_K$ corrections to the classical strong coupling are given by 
\be
\mathcal H_f^{(0,n,0)} = \sum_{ij}\tau^{(n)}_{ij}f^\dagger_i f_j \label{eq:H_0n0}
\ee
where $\tau^{(n)}_{ij}$ are effective hoppings describing perturbative corrections to the strong coupling limit of order $t^n/J_K^n$. These are determined in a straightforward manner from the transformed Hamiltonian and for $n=1$ and $n=2$ we find
\begin{align}
\tau^{(1)}_{ij} & = - \frac{1}{J_K}\sum_{k}t^{\up\down}_{ik}t^{\down\up}_{kj},  \label{eq:tau_ij(1)} \\
\tau^{(2)}_{ij} &=\frac{1}{2J_K^2}\sum_{kl}  \left(2 t^{\up\down}_{ik} t^{\down\down}_{kl } t^{\down\up}_{lj }-  t^{\up\up}_{ik }  t^{\up\down}_{kl} t^{\down\up}_{lj } -t^{\up\down}_{ik }  t^{\down\up}_{kl} t^{\up\up}_{lj } \right).  \label{eq:tau_ij(2)}
\end{align}
As is clear from the form of the effective hoppings, they are generated by virtual processes into the spin-down band~\cite{Koller:2002p144425}. 

In addition to perturbative hopping processes between sites the $t/J_K$ corrections also give rise to on-site terms, when $i=j$. In both cases, i.e., on-site and hopping terms, the perturbative corrections can be directly expressed in terms of the classical local moments, given by $\hat \bn_i$, which makes the dependence on the spin configuration transparent. In the case of the on-site terms, we find for the first order corrections
\be 
\tau^{(1)}_{ii} =  \frac{1}{2J_K}\sum_{k}t^2_{ik}(\hat \bn_i\cdot \hat \bn_k-1), \label{eq:tau_ii(1)-spins} 
\ee
and for the second order corrections
\be
\tau^{(2)}_{ii} = \frac{1}{4J_K^2}\sum_{kl}t_{ik}t_{kl} t_{li}(2 \hat \bn_k\cdot \hat \bn_l-\hat \bn_i\cdot \hat \bn_k-\hat \bn_i\cdot \hat \bn_l) .  \label{eq:tau_ii(2)-spins} 
\ee
These expressions highlight how the effective on-site energies are determined by the relative orientation of the classical local moments. In particular, whether or not such on-site energies arise is seen to depend on the type of spin configuration. For a ferromagnet the on-site corrections vanish, for instance, as expected. In general, one may distinguish two cases: the configuration spins either gives rise to a nonzero but uniform on-site energy or a nonuniform on-site energy. The latter is a more unusual scenario and implies that the magnetic texture induces a charge redistribution, which can be interpreted as a charge density wave. 

In a similar way, the first order effective hoppings between sites $i$ and $j$ can be expressed in terms of the classical spins as 
\begin{multline}
\tau^{(1)}_{ij} = - \frac{\Omega^{\up\up}_{ij}}{2J_K}\sum_{k}\frac{t_{ik}t_{kj} }{1+\hat \bn_i\cdot \hat \bn_j  } \left( 1+\hat \bn_i\cdot \hat \bn_j   \right. \\
\left.  - \hat \bn_i\cdot \hat \bn_k  -\hat \bn_j\cdot \hat \bn_k+ i \hat \bn_i \times \hat \bn_j  \cdot \hat \bn_k\right).\label{eq:tau_ij(1)-spins}
\end{multline}
This expression is valid as long as $\hat \bn_i\cdot \hat \bn_j \neq -1$, i.e., as long as the classical spins on sites $i$ and $j$ are not anti-aligned. A different expression applies in the case $\hat \bn_i\cdot \hat \bn_j = -1$.

The expressions in Eqs.~\eqref{eq:tau_ii(1)-spins}, \eqref{eq:tau_ii(2)-spins}, and \eqref{eq:tau_ij(1)-spins} demonstrate how a particular magnetic configuration affects the electrons which are strongly coupled to local moments via the Kondo coupling. Two effects in particular are noteworthy. As mentioned, one possibility is a nonuniform charge redistribution induced for special types of magnetic configurations, and a second possibility charge currents induced by magnetic textures for which $\hat \bn_i \times \hat \bn_j  \cdot \hat \bn_k$ is nonzero, as described by Eq.~\eqref{eq:tau_ij(1)-spins}. Such charge effects may be compared to charge effects which can occur in certain types of frustrated magnetic Mott insulators \cite{Bulaevskii:2008p024402}. In the case of the latter, it was pointed out that frustration can give rise to charge polarization and orbital currents in Mott insulators with particular magnetic ground states, and the expressions obtained for the polarization and orbital currents in terms of spin operators resemble the perturbative charge effects given by Eqs.~\eqref{eq:tau_ii(1)-spins}, \eqref{eq:tau_ii(2)-spins}, and \eqref{eq:tau_ij(1)}. From a symmetry perspective this is, of course, not surprising.

\subsection{Quantum effects on the fermion dynamics \label{ssec:f-quantum}}

Having the discussed the classical limit, we then proceed to a discussion of quantum effects, i.e., terms in the Hamiltonian originating from the treatment of spins as quantum degrees of freedom. In the notation of Eq.~\eqref{eq:H_mnl} these are terms with $m \ge 1$. 

We first consider corrections to the fermion Hamiltonian $\mathcal H_f$ which do not involve the bosons. At lowest order in $1/S$ ($m=1$ and $n=1$), these terms are given by
\be
\mathcal H^{(1,1,0)}_{f} =   \frac{1}{2J_K S}\sum_{ijk} t_{ik}^{\up\down}t_{kj}^{\down\up}(1 - f_{k}^{\dagger}f_{k}) f_{i}^{\dagger}f_{j}. \label{eq:deltaH-f}
\ee
It is clear that such terms only arise when $t_{ij}^{\up\down}$ is nonzero, and are therefore excluded in ferromagnets. Note that they do arise in collinear antiferromagnets. 

To understand the origin of such terms, it is instructive to consider the original Hamiltonian of Eq.~\eqref{eq:H_local} before canonical transformation (but after HP substitution). When viewed from the perspective of the original Hamiltonian, the form of Eq.~\eqref{eq:deltaH-f} can be understood as the result of a sequence of four processes described by~\eqref{eq:H_local}. First, a locally aligned spin-up electron can hop from site $j$ to site $k$ and flip its spin to locally anti-aligned (spin-down) at site $k$ in the process. Such a process creates an excitation with energy $\sim J_K$ and has matrix element $t_{kj}^{\down\up}$. At site $k$ a spin-flip process can then occur, whereby the spin is flipped and a spin-wave excitation is created. This is described by \eqref{eq:T^1/2_1}. The reverse process then removes the spin wave excitation and flips the spin back to anti-aligned (spin-down). Importantly, this on-site spin-flip process, which is enabled by the Kondo coupling to the local moment, can only occur if there is not already an aligned spin-up electron present at site $k$, which explains the factor $1 - f_{k}^{\dagger}f_{k}$ in~\eqref{eq:deltaH-f}. Note also that each on-site spin-flip process has matrix elements $\sim J_K/ \sqrt{S}$. Finally, another spin-flip hopping from site $k$ to site $i$ then completes the correlated hopping described by \eqref{eq:deltaH-f}. Note that the accounting of this sequence of processes yields the correct order of matrix elements in \eqref{eq:deltaH-f}, since $( t \times J^2_K/S \times t ) / J_K^3 = t^2 /J_K S$. 

Evidently, $\mathcal H^{(1,1,0)}_{f} $ in Eq.~\eqref{eq:deltaH-f} should be viewed as a quantum correction to $\mathcal H^{(0,1,0)}_{f} $, as given by Eqs.~\eqref{eq:H_0n0} and \eqref{eq:tau_ij(1)}, since both are of order $t^2 /J_K $. The correlated nature of this correction, captured by the factor $1 - f_{k}^{\dagger}f_{k}$, is a trace of the quantum nature of the spins. 

\subsection{Linear spin wave terms \label{ssec:LSW}}

Next, we consider the terms in the effective Hamiltonian which contribute to the linear spin wave theory of the magnon excitations. These comprise all boson terms up to and including order $1/S$ and thus correspond to $m=1$ in Eq.~\eqref{eq:H_mnl}. Terms of this kind come from both the Heisenberg coupling and the Kondo Hamiltonian. We first describe the strong coupling limit $n=l=0$ and then discuss the $t/J_K$ corrections ($n=1$).

The contribution from the Heisenberg coupling takes a form well-known from standard spin wave expansions and is given by
\begin{multline}
\mathcal H^{(1,0,0)}_{a} = \frac{1}{\sqrt{2S}}\sum_{ij} \left[ J_{ij} (\Gamma^{zx}_{ij}-i\Gamma^{zy}_{ij}) a_j  +\text{H.c.}) \right]   \\ 
+\frac{1}{2S} \sum_{ij}J_{ij}  \left[  ( A_{ij} a^\dagger_ia_j  +B_{ij} a^\dagger_ia^\dagger_j +\text{H.c.} ) \right. \\
\left. - \Gamma^{zz}_{ij} (n_i+n_j)\right] , \label{eq:H_a}
\end{multline}
where $\Gamma^{ab}_{ij}$, $A_{ij}$, and $B_{ij}$ were defined in Eqs.~\eqref{eq:Gamma} and \eqref{eq:A_ijB_ij}. (See also Appendix~\ref{app:HP-J}.) It is important to emphasize that here we have kept terms linear in the bosons, which are of order $S^{-1/2}$. Such terms do not arise in pure Heisenberg magnets, since spin wave expansions are performed around the classical ground state, i.e., the state which minimizes the classical energy. By definition, linear terms vanish when expanding around a minimum. Here these terms do not necessarily vanish, since the classical ground state need not minimize the classical Heisenberg energy, but instead may arise from a competition of the Heisenberg coupling and the Kondo Hamiltonian, and thus minimize the total classical energy. 

In the strong coupling limit the contribution from the Kondo Hamiltonianis given by
\begin{multline}
\mathcal H^{(1,0,0)}_{f,a} = \frac{1}{\sqrt{2S}} \sum_{ij} (t^{\up\down}_{ij} a_j +t^{\down\up}_{ij} a^\dagger_i ) f^\dagger_i f_j \\
+ \frac{1}{4S}\sum_{ij} (2 t^{\down\down}_{ij} a^\dagger_ia_j -t^{\up\up}_{ij} a^\dagger_i a_i-t^{\up\up}_{ij} a^\dagger_j a_j) f^\dagger_i f_j ,  \label{eq:H_fa}
\end{multline} 
which shows that its impact on the boson dynamics intrinsically depends on the fermions. Indeed, without a Heisenberg contribution the bosons have no bare dispersion. Note that $\mathcal H_{f,a} $ also has terms linear in the bosons, as does $\mathcal H_{a}$. As mentioned, this reflects the fact that, in general, the classical ground state minimizes the total classical energy, but not necessarily each individual contribution to the total energy. While the linear terms can survive in the Hamiltonian, the sum of all linear terms does vanish after taking the average over the fermions. The linear terms therefore do not contribute to the boson dispersion at first order, which restores the notion that linear terms vanish when expanding around a minimum. Note that the linear terms may contribute at second order, however, yielding a contribution to the magnon dispersion of order $1/S$. 

To determine the linear spin wave spectrum beyond the strong coupling limit, one must include the $t/J_K$ corrections. In the notation of Eq.~\eqref{eq:H_mnl}, this means including $n=1$ (or even higher) terms. Collecting all relevant corrections using the recipe of Sec.~\ref{ssec:H-expand} is straightforward but quickly becomes tedious. In particular, collecting all corrections for the most general ground state configuration even to first order in $t/J_K$ yields a lengthy contribution to $ \mathcal H^{(1,1,0)}_{f,a}$. The form of this contribution simplifies, however, under the assumption that $t^{\up\down}_{ij}=t^{\down\up}_{ij}=0$, in which case the first order $t/J_K$ corrections take the form
\begin{multline}
 \mathcal H^{(1,1,0)}_{f,a} = \frac{1}{4S J_K }  \sum_{ijk}  \left[ 2t^{\up\up}_{ik} t^{\down\down}_{kj} a^\dagger_ka_j + 2t^{\down\down}_{ik} t^{\up\up}_{kj} a^\dagger_ia_k \right.  \\
\left. - 2  t^{\down\down}_{ik} t^{\down\down}_{kj} a^\dagger_i a_j - t^{\up\up}_{ik} t^{\up\up}_{kj}(a^\dagger_ia_i +a^\dagger_j a_j ) \right]  f^\dagger_{i}f_{j} \label{eq:tJ_K-correct}
\end{multline}
The form of \eqref{eq:tJ_K-correct} further simplifies in the case of a ferromagnetic ground state configuration, in which case all $t^{\sigma\sigma}_{ij}  $ can be replaced by $t_{ij}$.

\subsection{Quantum corrections to linear spin wave terms \label{ssec:a-quantum}}

Next, we consider contributions to the effective Hamiltonian which represent quantum corrections to the linear spin terms, i.e., terms up to order $1/S^2$. Here we focus in particular on the strong coupling limit, which corresponds to $n=l=0$. As in the case of linear spin wave order, there are two sources for such terms: the Heisenberg coupling and the Kondo coupling. The contributions from the former are obtained by carrying out the familiar spin wave expansion of a Heisenberg model to the next order in $1/S$. Going beyond linear spin wave order is less common in the context of magnetism described by Heisenberg models, but is nonetheless a standard affair and we will therefore not discuss these contributions here in detail. We refer to Appendix~\ref{app:HP-J} for an expansion of the Heisenberg coupling up to order $1/S^2$.

Here we instead focus on the contributions from the Kondo coupling, or, more precisely, on the contributions which would not arise without the Kondo coupling. These are collected in $\mathcal H^{(2,0,0)}_{f,a}$. It is in principle straightforward to compute all terms of the desired order using the formalism of Sec.~\ref{ssec:H-expand}. As mentioned above, for a general magnetic ground state this is tedious and the resulting expressions are lengthy. We therefore only quote the result for a ferromagnetic ground state, in which case $t^{\sigma\sigma}_{ij}  \rightarrow t_{ij}$ and $\Gamma^{ab}_{ij} \rightarrow \delta^{ab}$. For a ferromagnetic ground state we find
\begin{multline}
\mathcal H^{(2,0,0)}_{f,a} =   \frac{1}{32S^2}\sum_{ij} t_{ij} \left[3(a_i^{\dagger}a_i + a_j^{\dagger}a_j)-8  a_i^{\dagger}a_j \right]f_{i}^{\dagger}f_{j} \\
 - \frac{1}{32S^2}\sum_{ij}t_{ij}\left[a_i^{\dagger}a_i^{\dagger}a_ia_i - a_i^{\dagger}a_ia_j^{\dagger}a_j + (i\leftrightarrow j)\right]f_{i}^{\dagger}f_{j} \\
  + \frac{1}{4S^2}\sum_{ij}J_{ij}(2a_i^{\dagger}a_i    - a_i^{\dagger}a_j - a_j^{\dagger}a_i) f^\dagger_{i}f_{i}. \label{eq:H_fa-QC}
\end{multline}
The first two terms are in full agreement with Ref.~\onlinecite{Shannon:2002p104418} and can be used to compute quantum corrections to the spin wave dispersion of a double-exchange ferromagnet. The third term represents a contribution from the Heisenberg coupling, since it is proportional to $J_{ij}$, but its presence relies on the Kondo coupling. The latter follows from the fact it describes an interaction between fermions and bosons. This term therefore shows that the quantum nature of the local moments spins generates an interaction between the Kondo and Heisenberg couplings and gives rise subtle quantum effects affecting the spin wave dynamics.


\subsection{Fermions in a local spin $S-1/2$ state \label{ssec:f-down}}

In the final part of this section we turn to a brief discussion of the effective Hamiltonian for the transformed spin-down electrons, given by $\tilde f_{i\down}$ in Eq.~\eqref{eq:f_down-transform}. As discussed in Sec.~\ref{sec:discussion}, the transformed spin-down electrons should be interpreted as electrons in a state of spin $S-1/2$ with the local moment spin. Since such a state is quantum mechanically distinct from an electron in a $S+1/2$ spin state, the effective Hamiltonian will reflect this difference. The effective Hamiltonian for the transformed spin-down electrons is of interest when the Kondo coupling is antiferromagnetic, such that the $S-1/2$ spin states form the low-energy degrees of freedom. 

As for the transformed spin-up electrons, we simplify the notation by denoting the transformed spin-down electrons operators $d_i$. The effective Hamiltonian can then be expressed in a way analogous to the $f$-electrons. Now, $\mathcal H_d$ collects all terms which only depend on the fermions and it is here where the differences with the $f$-electron are most clearly reflected. Using notation similar to Eq.~\eqref{eq:H_mnl}, the lowest order contributions are given by
\be
\mathcal H_d = \mathcal H^{(0,0,0)}_{d} +\mathcal H^{(0,1,0)}_{d}+\mathcal H^{(1,0,0)}_{d} +\mathcal H^{(1,1,0)}_{d}.
\ee
The first two terms on the right hand side are direct analogs of \eqref{eq:H_DE} and \eqref{eq:tau_ij(1)} but with $\up $ and $ \down$ exchanged. The third and fourth term arise only in the case of $d$-electrons and are absent for $f$-electrons. Explicitly, the third term is given by
\be
\mathcal H^{(1,0,0)}_{d} =  - \frac{1}{2S} \sum_{ij} t^{\down\down}_{ij} d^\dagger_id_j,
\ee
and the fourth term takes the form
\be
\mathcal H^{(1,1,0)}_{d} =  \frac{1}{2SJ_K}\sum_{ijk} \left[t^{\down\down}_{ik} t^{\down\down}_{kj}  + (t^{\up\up}_{ik} t^{\up\up}_{kj} - t^{\up\down}_{ik} t^{\down\up}_{kj}) \delta_{ij}\right]d^\dagger_i d_j.
\ee
Both terms are bilinear in the fermions and are of order $1/S$. As a result of the latter, both terms represent quantum corrections to the classical limit arising from the quantum nature of the local moment spin. They are present for the $S-1/2$ electron states due to the intrinsic quantum nature of these states, as discussed in Sec.~\ref{sec:klm} [see Eq.~\eqref{eq:psi_-}]. These terms imply that, even at the non-interacting level, the dispersion of the $d$-electrons is affected by the quantum nature of the local moment spins. 


\section{Generalization: spin-orbit coupling  \label{sec:soc}}

As mentioned in the introduction, there are two possible generalizations of the canonical spin wave expansion. These generalizations are presented in this section and in the next. The first is natural and straightforward, and is achieved by including the effect of spin-orbit coupling. In the presence of spin-orbit coupling, the spin degrees of freedom---both of the electrons and the local moments---are locked to the lattice, such that spin and spatial degrees of freedom transform jointly under the symmetries of the crystal lattice. A separate full spin rotation symmetry is absent. As a result, both the hopping part of the Kondo Hamiltonian and the exchange coupling between the spins take a more general form. (Here we do not consider modifications of the Kondo coupling.)

\subsection{Modifications with spin-orbit coupling}

In the case of the hopping Hamiltonian, which we denote $H_t$, the general form in the presence of spin-orbit coupling is given by
\be
H_t = \sum_{ij} c^\dagger_{i}  h_{ij}  c_{j},
\ee 
where $ h_{ij}  $ is now a matrix in spin space. Hopping processes may now depend on spin. It is convenient to expand the hopping matrix $ h_{ij} $ in terms of Pauli spin matrices as
\be
h_{ij}  = t_{ij} + i \blambda_{ij}\cdot \bsigma. \label{eq:t_ij_soc}
\ee 
where $t_{ij}$ is the uniform spin-independent hopping considered up to this point, and $\blambda_{ij}=-\blambda_{ji}$ are real and describe the spin-dependent hopping introduced by spin-orbit coupling. When $\blambda_{ij}=0 $ the hopping Hamiltonian reduces to the form of~\eqref{eq:H_klm}. 

Spin-orbit coupling also affects the Heisenberg coupling, which we denote $H_J$. The lack of spin rotation invariance can give rise to anisotropic exchange couplings, such that the most general bilinear coupling of pairs of spins in the presence of spin-orbit coupling is given by
\be
H_J  = \frac{1}{2S^2} \sum_{ij} J_{ij}^{ab} S^a_i S^b_j. \label{eq:H_J_soc}
\ee 
The exchange couplings $J_{ij}^{ab}$ now have matrix structure in spin space and the form the exchange couplings for a given pair of sites $(ij)$ is determined by the crystal symmetries of the system in question. Full rotational invariance in spin space would require $J_{ij}^{ab}=J_{ij}\delta^{ab}$ and this indeed recovers the form of Eq.~\eqref{eq:H_klm}. 

These generalizations of the hopping and the exchange couplings are readily incorporated in the formalism of the canonical spin wave expansion. In fact, the only change occurs at the level of $t^{\sigma\sigma'}_{ij}$ defined in Eq.~\eqref{eq:t_ij_ss}, and $\Gamma_{ij} $ defined in Eq.~\eqref{eq:Gamma_ij}. In the presence of spin-orbit coupling $t^{\sigma\sigma'}_{ij}$ takes the form
\be
t^{\sigma\sigma'}_{ij} = (U^\dagger_i h_{ij} U_j)_{\sigma\sigma'}, \label{eq:t_ij_local_soc}
\ee
and $\Gamma_{ij} $ becomes
\be
\Gamma_{ij}  = R^T_i J_{ij} R_j. \label{eq:Gamma_ij_soc}
\ee 
Where previously hopping and exchange anisotropies only originated from the site-dependent local frames associated with classical spin configuration, they now also have an intrinsic component originating from spin-orbit coupling. And while the explicit expressions for $t^{\sigma\sigma'}_{ij}$ and $\Gamma_{ij}$ are different, the structure and formalism of the spin wave expansion remain unaltered. 

The final form of the effective Hamiltonian does change, however, and will now depend on $\blambda_{ij}$, as well as the exchange anisotropies of $J^{ab}_{ij}$ introduced by spin-orbit coupling. To examine some of the implications of these anisotropies, it is most insightful to consider how $\blambda_{ij}$ affects the classical limit of Kondo lattice model, which was discussed for the isotropic case in Sec.~\ref{ssec:classical}. 


\subsection{Effective magnetic anisotropy}

We first focus on the effective hopping of low-energy electrons in the strong coupling limit, which is described by Eq.~\eqref{eq:H_DE}, and determine how $t^{\up\up}_{ij}$ changes in the presence of spin-orbit coupling. We are interested in particular in the amplitude of the effective hopping (i.e. absolute value), since it is the amplitude which, in the isotropic case, captures the ferromagnetic tendency of the Kondo lattice model in the strong coupling double-exchange limit. Indeed, as first pointed out by Anderson and Hasegawa~\cite{Anderson:1955p675}, in the isotropic case one has
\be
|t^{\up\up}_{ij}|  = t_{ij} \sqrt{ \tfrac12 (1+ \hat \bn_i\cdot \hat \bn_j)}, \label{eq:abs_tij}
\ee
which is maximal for ferromagnetic alignment of $ \bn_i$ and $\hat \bn_j$. Since the magnetic ground state seeks to minimize the electron kinetic energy, which is achieved by maximizing the electron bandwidth, a ferromagnetic ground state is strongly favored. 

In the presence of spin-orbit coupling it is straightforward to show that $|t^{\up\up}_{ij}|^2 = \tr{P^+_i h_{ij}P^+_j h_{ji}}$, which reduces to
\begin{multline}
2|t^{\up\up}_{ij}|^2 = t^2_{ij} ( 1+ \hat \bn_i\cdot \hat \bn_j)  + \blambda^2_{ij} ( 1- \hat \bn_i\cdot \hat \bn_j) \\
 +2 ( \hat \bn_i\cdot \blambda_{ij} )(  \hat \bn_j \cdot\blambda_{ij} ) + 2 t_{ij }\blambda_{ij}\cdot  \hat \bn_i \times \hat\bn_j , \label{eq:abs_tij_SOC}
\end{multline}
such that the effective hopping follows by taking the square root. It is clear from Eq.~\eqref{eq:abs_tij_SOC} that spin-orbit coupling introduces bond-dependent magnetic anisotropies tied to the direction of $\blambda_{ij}$, and thus in general introduces competing ordering tendencies. To examine the effect of $\blambda_{ij}$ in more detail, it is useful to consider the specific case where only $\lambda^z_{ij}$ is nonzero. The general case follows readily from this special case. When $\blambda_{ij}$ points in the $z$ direction Eq.~\eqref{eq:abs_tij_SOC} simplifies and reduces to (suppressing subscripts and setting $\lambda = \lambda^z$)
\begin{multline}
2|t^{\up\up}|^2  = ( t^2 +\lambda^2) ( 1+ \hat n^z_i  \hat n^z_j )  \\
+ \frac12 \left[ ( t + i\lambda)^2 \hat n^+_i  \hat n^-_j + \text{c.c.}\right] ,    \label{eq:abs_tij_SOC-2}
\end{multline}
where $\hat n^\pm _i  =  \hat n^x_i \pm i  \hat n^y_i$ (and similarly for $j$). A further important simplification is achieved by making the substitution 
\be
\hat n^+_i \rightarrow e^{- i\gamma }\hat n^+_i , \quad  \hat n^+_j \rightarrow e^{i\gamma }\hat n^+_j, \label{eq:n-rotate}
\ee
which should be interpreted as a rotation of $\hat \bn_i$ around the $z$ axis by an angle $\gamma = \tan^{-1}(\lambda/t)$, and a similar rotation of $\hat \bn_j$ in the opposite direction. Denoting the rotated classical spin vectors on site $i$ and $j$ as  $\hat \bn'_i $ and $ \hat \bn'_j$, the absolute value of the effective hopping can expressed as
\be
|t^{\up\up}_{ij}|  =\sqrt{ \tfrac12 (t^2_{ij} +\lambda^2_{ij}) (1+ \hat \bn'_i\cdot \hat \bn'_j)},  \label{eq:abs_tij_SOC-rot}
\ee
which is essentially equivalent to the result of the isotropic case. It is easy to see that \eqref{eq:abs_tij_SOC-rot} also holds in the general case, when $\blambda_{ij}$ is not chosen along $z$, as long as the rotation defined in \eqref{eq:n-rotate} is performed around the direction of $\blambda_{ij}$.

\begin{figure}
	\includegraphics[width=0.9\columnwidth]{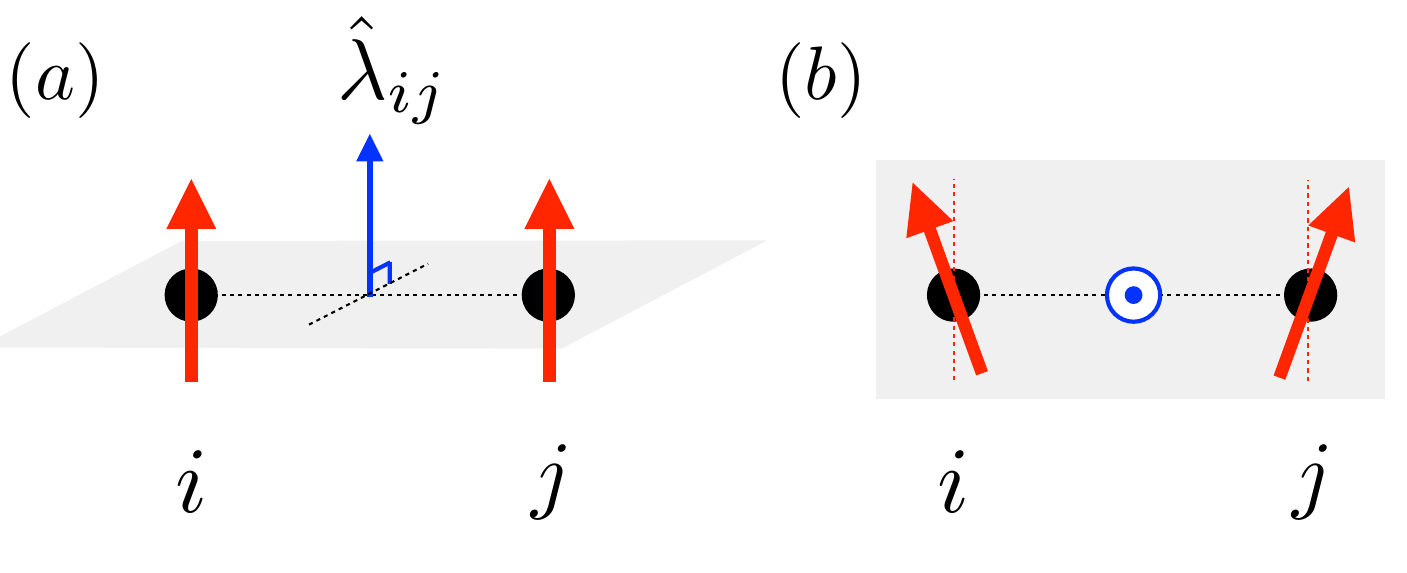}
	\caption{Sketch of the spin orientations which maximize the hopping amplitude $|t^{\up\up}_{ij}|$ in the presence of spin-orbit coupling. Either the spins on sites $i$ and $j$ are aligned along $\blambda_{ij}$, shown in (a), or the spins lie in the plane normal to $\blambda_{ij}$ and make angle $2\gamma$ (see text), shown in (b).}
	\label{fig:sketch_soc}
\end{figure}

A number of conclusions now follow. First, Eq.~\ref{eq:abs_tij_SOC-rot} suggests that minimizing the electron kinetic energy is now favored by a ferromagnetic alignment of $\hat \bn'_i $ and $\hat \bn'_j$, i.e., ferromagnetic alignment of spins in the rotated spins. ``Ferromagnetic'' alignment should therefore be understood in reference to a spin frame determined by $\blambda_{ij}$, and is therefore in general bond-dependent. In particular, it follows from \eqref{eq:abs_tij_SOC-rot} that the effective hopping amplitude is maximized when (the unrotated spins) $\hat \bn_i $ and $\hat \bn_j$ are ferromagnetically aligned in the direction of $\blambda_{ij}$, or lie in the plane perpendicular to $\blambda_{ij}$ and make an angle $2\gamma_{ij}$. This is shown schematically in Fig.~\ref{fig:sketch_soc}(a) and \ref{fig:sketch_soc}(b), respectively. Therefore, as one might expect, spin-orbit coupling introduces easy-axis and easy-plane anisotropies. 

Spin-orbit coupling furthermore introduces frustration, since $\blambda_{ij}$ may point in different directions depending on the bond pair $(ij)$, and it is not guaranteed that ferromagnetic alignment in each bond is compatible. Hence, which ordering pattern is most favorable, that of Fig.~\ref{fig:sketch_soc}(a) or Fig.~\ref{fig:sketch_soc}(b), can only be determined by minimization of the energy of the lattice system, and is thus model and lattice dependent.

Next, we consider the hopping matrix elements $t^{\up\down}_{ij}$ connecting the spin-up and spin-down bands. These enter in the $t/J_K$ corrections given by Eqs.~\eqref{eq:tau_ij(1)}, which in the isotropic case reduces to \eqref{eq:tau_ii(1)-spins} for the on-site corrections.  To compare with this isotropic result, we determine the form of $|t^{\up\down}_{ij}|^2 = \tr{P^+_i h_{ij}P^-_j h_{ji}}$ and find
\begin{multline}
2|t^{\up\down}_{ij}|^2 = t^2_{ij} ( 1- \hat \bn_i\cdot \hat \bn_j)  + \blambda^2_{ij} ( 1+ \hat \bn_i\cdot \hat \bn_j) \\
 -2 ( \hat \bn_i\cdot \blambda_{ij} )(  \hat \bn_j \cdot\blambda_{ij} ) - 2 t_{ij }\blambda_{ij}\cdot  \hat \bn_i \times \hat\bn_j .\label{eq:abs_tij_SOC_updown}
 \end{multline}
This expression has a structure similar to \eqref{eq:abs_tij_SOC} and also emphasizes the magnetic anisotropy introduced by spin-orbit coupling. Performing similar manipulations as in \eqref{eq:abs_tij_SOC-2} and \eqref{eq:n-rotate}, we can bring this expression into the form
 \be
|t^{\up\down}_{ij}|^2 = \frac12 (t^2_{ij} +\lambda^2_{ij}) (1- \hat \bn'_i\cdot \hat \bn'_j)  ,\label{eq:SOC_t_ij_updown}
 \ee
 which again resembles the isotropic result, but for rotated spins $ \hat \bn'_i $ and $ \hat \bn'_j$. Comparing with \eqref{eq:tau_ii(1)-spins}, we conclude that \eqref{eq:SOC_t_ij_updown} implies a perturbatively generated antiferromagnetic coupling between $ \hat \bn'_i $ and $ \hat \bn'_j$. Since the spin are rotated, however, this can lead to noncollinear spin arrangements in the physical unrotated basis.

The role of spin-orbit coupling in itinerant Kondo lattice magnets has recently come into focus as a possible mechanism for stabilizing unconventional and exotic spin textures such as Skyrmion crystals. The effect of spin-orbit coupling on the magnetic phase diagram of Kondo lattice magnets has been examined both in the weak coupling RKKY limit~\cite{Hayami:2018p137202,Okada:2018p224406} and in the strong coupling double-exchange limit~\cite{Banerjee:2014p031045,Meza:2014p085107,Zhang:2020p024420,Kathyat:2020p075106,Kathyat:2021p035111}. In particular, when applied to metals with Rashba spin-orbit coupling, Eq.~\eqref{eq:abs_tij_SOC} reduces to the model considered in Ref.~\onlinecite{Kathyat:2020p075106}. Similarly, Ref.~\onlinecite{Mukherjee:2021p134424} studied a Rashba double-exchange model at half filling, i.e., at a density of one electron per site, in which case the band of spin-aligned fermions is completely filled. In that case the charge is effectively frozen and the interactions between the local moments are determined by the perturbative coupling to the spin-anti-aligned band, which are of order $t^2 /J_K$ and given by Eqs.~\eqref{eq:SOC_t_ij_updown} and \eqref{eq:SOC_t_ij_updown}. The model obtained in Ref.~\onlinecite{Mukherjee:2021p134424} for the Rashba case is indeed a particular case of Eq.~\eqref{eq:SOC_t_ij_updown}. Note further that while here we have focused attention on the hopping amplitude, another important question is how Berry curvature effects, stemming from the hopping phases, are affected by spin-orbit coupling. This question was addressed in detail in Ref.~\onlinecite{Zhang:2020p024420}. 

It will be interesting to fully explore the similarities and differences between spin-orbit coupled metallic magnets and Heisenberg magnets with exchange anisotropy. One unexplored but fruitful direction is to compare the low-energy magnetic excitations. The generalization of the canonical spin wave expansion to spin-orbit coupled Kondo lattice magnets will enable the study of the spin wave spectrum in these systems.


\section{Generalization: magnetic superconductors  \label{sec:SC}}

The second generalization we propose aims to include superconductivity. This generalization is more consequential and requires a modification of the structure of the spin wave expansion itself. 

To include the effect of $s$-wave electron pairing in a BCS mean-field setting, we add to the Kondo-Heisenberg Hamiltonian of Eq.~\eqref{eq:H_klm} an on-site spin-singlet pairing Hamiltonian $H_\Delta$ given by
\be
H_\Delta= \sum_{i} \Delta_0 c^\dagger_{i\up}c^\dagger_{i\down} +\text{H.c.}. \label{H_Delta}
\ee
The full Hamiltonian then describes systems in which superconductivity (of electrons) coexists with local moment spin degrees of freedom. More specifically, our focus here is on systems which simultaneously realize superconductivity and magnetism. These may be intrinsic magnetic superconductors, often exhibiting unconventional helical magnetism, or hybrid heterostructure systems relying on the superconducting proximity effect.

A particularly prominent example of the latter, which has attracted much attention, are effectively one-dimensional (1D) arrays of magnetic adatoms deposited on superconducting substrate surfaces. In such systems, when the magnetic moments develop the right type of magnetic order---for instance spiral order---the resulting phase is a 1D topological superconductor equivalent to a Kitaev wire~\cite{Choy:2011p195442,Martin:2012p144505,Nadj-Perge:2013p020407,Klinovaja:2013p186805,Pientka:2013p155420,Vazifeh:2013p206802,Braunecker:2013p147202}. As as result, systems of this kind, combining proximitized superconductivity and magnetism, have become a versatile platform for seeking realizations of topological superconductivity and Majorana bound states~\cite{Nadj-Perge:2014p602}. The Kondo lattice model of Eq.~\eqref{eq:H_klm} together with $H_\Delta$ provide a good description for this class of systems, both in 1D and in 2D~\cite{Nakosai:2013p180503,Li:2016p12297,Soldini:2023p1848}.


\subsection{Generalization of the canonical spin wave expansion}

Our goal in this section is to extend the canonical spin wave expansion to magnetic superconductors. This is achieved by applying the formalism developed in Sec.~\ref{sec:transform} to the Kondo lattice Hamiltonian with additional pairing term \eqref{H_Delta}. Here we describe what modifications of the formalism are necessary and what the resulting effective Hamiltonian is. Recall that the spin wave expansion scheme involves the following basic steps: (i) transforming to spin frame tied to the magnetically ordered state, (ii) writing the local moment spins in terms of HP bosons, and (iii) performing a strong coupling expansion. 

Since, by assumption, the pairing is conventional spin-singlet $s$-wave pairing, the pairing term is unaffected by the local unitary rotation of Eq.~\eqref{eq:f-fermions}. As a result, after rotating to the local basis the pairing term takes the form $H_\Delta = \Delta + \Delta^\dagger$, with $\Delta$ given by
\be
\Delta= \sum_{i} \Delta_0 f^\dagger_{i\up}f^\dagger_{i\down}.
\ee
After substituting the HP boson representation, the full Hamiltonian is then written as 
\be
H = H_0 + T_0 + T_1 + T_{-1}+\Delta+\Delta^\dagger, \label{Ham}
\ee
where the $T_m$ operators are defined as before in Sec.~\ref{ssec:HPboson}. Equation \eqref{Ham} is the generalization of Eq.~\eqref{eq:H_Tm}. The term $H_0$ is still given by \eqref{eq:H_0}, except that now we must explicitly include the chemical potential $\mu$. The energies $\varepsilon_{\up,\down}$ are thus given by $\varepsilon_{\up,\down} = \mp J_K/2 -\mu$.
In addition to the commutators $[T_m, H_0]$ of Eq.~\eqref{eq:commutator}, which are unaltered by the chemical potential (hence its suppression up to now), we also require the commutator of $\Delta$ with $H_0$; we find
\be
[\Delta, H_0] = 2\mu \Delta, \quad [\Delta^\dagger, H_0] = -2\mu \Delta^\dagger. \label{commutator-Delta}
\ee
Having determined these commutators, we can proceed and derive a canonical transformation of the Hamiltonian \eqref{Ham} following the same approach used in Secs.~\ref{ssec:derivation}. The steps taken to derive the form of $e^{iQ}$ are fully analogous; the only modification comes from the pairing terms. Recall that the canonical transformation is defined and constructed by requiring that all spin-flip terms are removed up to the desired order. Since $\Delta$ and $\Delta^\dagger$ create or annihilate pairs of electrons of opposite spin, i.e., pairs of high- and low-energy fermions, we similarly seek to remove these terms from the Hamiltonian. As before, this requirement determines the form of $Q^{(p)}$ and of $\mathcal H^{(p)}$. Consider, for instance, the first order contribution $Q^{(1)}$ to $Q$, previously given by Eqs.~\eqref{eq:Q1-def} and  \eqref{eq:Q1}. Its modified form in the presence of pairing is defined by the condition 
\be
 \left[iQ^{(1)}, H_0\right] +T_1 + T_{-1} +\Delta+\Delta^\dagger= 0 ,
\ee
which can be solved using the commutators and yields
\be
Q^{(1)} = i(T_1 - T_{-1} )/J_K+ i(\Delta-\Delta^\dagger)/2\mu.
\ee
The form of $Q^{(1)}$, in turn, determines the form $\mathcal H^{(2)}$, which is now given by 
\begin{multline}
\mathcal H^{(2)} = -\frac{1}{J_K} \left[T_1, T_{-1} \right]- \frac{1}{2\mu}\left[\Delta ,\Delta^\dagger \right] 
 - \\
  \left\{ \left(\frac{1}{J_K}-\frac{1}{2\mu} \right)[T_1, \Delta] +\left(\frac{1}{J_K}+\frac{1}{2\mu} \right)[T_1, \Delta^\dagger ] \right. \\
\left.  + \text{H.c.} \right\}. 
\end{multline}
Comparing this expression to Eq.~\eqref{eq:H(2)} shows that additional terms arise in the transformed Hamiltonian due the pairing term $\Delta$. A subset of these terms constitute effective pairing terms. Indeed, the canonical transformation generates perturbative terms which correspond to pairing of electrons of the same spin, but on different sites. As a result, such terms describe Cooper pairs with nonzero angular momentum. The precise form of such pairing terms depends on the underlying magnetic state.

\subsection{Effective Hamiltonian in the presence of pairing}

We briefly consider the form of the effective Hamiltonian in the presence pairing. The effective Hamiltonian is derived in the same manner as described in Sec.~\ref{sec:H_eff}. Here we do not give a full systematic account, but focus on differences that arise in the presence of pairing to lowest order. As in  Sec.~\ref{sec:H_eff} we consider the case of the $S+1/2$ $f$-fermions. 

The Hamiltonian for the $f$-fermions $\mathcal H_f $ takes the form
\be
\mathcal H_f =  \sum_{ij} \tau_{ij} f^\dagger_i f_j - \tilde \mu \sum_i f^\dagger_i f_i  + \sum_{ij} ( \tilde\Delta_{ij} f^\dagger_i f^\dagger_j +\text{H.c.} ) \label{eq:H_f_Delta}
\ee
with effective hopping $\tau_{ij}$ defined as
\be
\tau_{ij}=  t^{\up\up}_{ij} -\frac{1}{J_K}\sum_k  t^{\up\down}_{ik}  t^{\down\up}_{kj} . 
\ee
The effective chemical potential $\tilde \mu $ and, most importantly, the effective pairing $\tilde\Delta_{ij}$ are given by
\be
\tilde \mu =  \mu\left(1 + \frac{|\Delta_0|^2}{2\mu^2}\right), \;\; \tilde\Delta_{ij} =  \Delta_0\left(\frac{ 1}{2J_K} - \frac{1 }{4\mu}\right)  t^{\up\down}_{ij} . \label{eq:Delta_ij_tilde}
\ee
It is clear from this expression that the effective pairing depends on the magnetic state via $t^{\up\down}_{ij} \sim \bz^\dagger_i \bw_j$, and would vanish in the specific case of a ferromagnet. A nonzero pairing for the $f$-fermions is only induced for non-ferromagnetic magnetic states in which the spins are not fully aligned. This was first noted in Ref.~\onlinecite{Choy:2011p195442}, which pointed out that in the particular case of a single-$Q$ spiral state the effective pairing Hamiltonian is equivalent to the Kitaev wire model. 

The Hamiltonian of Eq.~\eqref{eq:H_f_Delta} is the strong coupling expansion of a magnetic superconductor when the spins are treated classically, and in that sense---as before---represents the classical limit. The spin wave piece of the effective Hamiltonian, given by $\mathcal H_{f,a}$, is also affected by the pairing. We can write $\mathcal H_{f,a} $ as a sum of two terms,
\be
\mathcal H_{f,a}  = \mathcal H^t_{f,a}  +\mathcal H^\Delta_{f,a}  ,
\ee
where the first corresponds to contributions already determined and given (to lowest order) by Eq.~\eqref{eq:H_fa}, and the second term collects additional contributions in the presence of pairing. These contributions are given by
\begin{multline} 
\mathcal H^\Delta_{f,a}  = \frac{1}{\sqrt{8S}}\left(\frac{J_K}{4\mu^2} + \frac{1}{J_K} \right)  \\
\times \sum_{ij}\left[\Delta_0\left(t_{ij}^{\down\down}a_i^{\dagger}- t_{ij}^{\up\up}a_j^{\dagger} \right)f_{i}^{\dagger}f_{j}^{\dagger}   +\text{H.c.} \right] \\
  + \frac{1}{4J_KS}\sum_{ij} \left(a_i^{\dagger}a_i + a_j^{\dagger}a_j \right)\left(\Delta_0t_{ij}^{\up\down}f_{i}^{\dagger}f_{j}^{\dagger} + \text{H.c.} \right),  \label{eq:H_fa_Delta}
\end{multline} 
showing that the spin wave dispersion it linear order will be affected by pairing. Note that these contributions are smaller by a factor of $\Delta_0 / J_K$ compared to \eqref{eq:H_fa}.


\section{Discussion and Conclusion \label{sec:conclusion}}

The central result of this paper is the development of a systematic strong coupling spin wave expansion for itinerant Kondo lattice magnets. This spin wave expansion relies on a canonical transformation of the Kondo lattice Hamiltonian which eliminates the coupling between the high- and low-energy sectors of the strong Kondo coupling regime. The resulting effective Hamiltonian describes the low-energy dynamics of effectively spinless electrons and spin wave bosons, which may be determined to any desired order in $1/S$ and $t/J_K$, as well as $J/J_K$ in cases where a Heisenberg coupling between the local moments is included. In the classical limit, given by the zeroth order in $1/S$, the effective Hamiltonian is simply a strong coupling expansion of the classical Kondo lattice Hamiltonian. 

A key feature of the canonical strong coupling spin wave expansion is the nature of the transformed fermion and boson degrees of freedom. As demonstrated in detail in Sec.~\ref{sec:discussion}, the transformed fermion operators correspond to electrons in a state of total spin $S \pm 1/2$ with the local moment spin, which are eigenstates of the on-site Kondo exchange coupling and are manifestly different from the original spin-up and spin-down states. The transformed boson operators correspond to spin wave excitations of the total spin on a given site, i.e., the total spin with quantum number $S \pm 1/2$. These degrees of freedom are natural in the (very) strong Kondo coupling regime, where it is appropriate to choose a basis in which the Kondo exchange coupling is diagonal. The canonical transformation which produces the spin wave expansion can thus be interpreted as a transformation to an eigenbasis of the Kondo coupling, determined up to the desired order in $1/S$ and $t/J_K$. It is important to emphasize that the new transformed fermions, as well as the bosons, are non-local operators in terms of the untransformed original electrons and bosons. Moreover, the transformed electrons can be shown to describe spin polaron states, i.e., electrons dressed with spin flip excitations, which reveals that the spin wave expansion formalism naturally describes spin polaron formation, a phenomenon known to occur in Kondo magnets. 

The spin wave expansion is constructed such that it applies to general magnetically ordered states and not just to ferromagnets. It is therefore applicable, for instance, to the tetrahedral spin-chiral order on the triangular lattice, predicted first in Ref.~\onlinecite{Martin:2008p156402} (and subsequently studied in the strong coupling regime in Ref.~\onlinecite{Kumar:2010p216405}), and recently reportedly observed in experiment~\cite{Park:2023p8346}. Our formalism stands to shed important light on the nature of the magnetic excitations in such non-coplanar itinerant magnets, complementing previous work in the weak coupling regime~\cite{Akagi:2013p123709}. 

In the companion paper to this paper we apply the spin wave expansion to realizations of Kondo lattice models on various lattices and study two aspects in particular: $t/J_K$ corrections to the spin wave dispersion of the strong coupling limit and the effect of spin-orbit coupling on the spin wave dispersion. Both questions are readily addressed using the spin wave formalism developed here and give important insight into the differences and similarities of itinerant Kondo and Heisenberg magnets. It is known, for instance, that the leading order spin wave dispersion of a Kondo lattice ferromagnet in the strict strong coupling limit is identical in form to the that of a Heisenberg ferromagnet. Furthermore, spin-orbit coupling generally introduces anisotropies in Heisenberg magnets, which is reflected in the spin wave dispersion (e.g. gapped magnetic excitations). The companion paper will examine in detail how $t/J_K$ corrections and spin-orbit coupling are revealed in the spin wave dispersion of itinerant Kondo magnets. 

To further elucidate the nature and adequacy of the canonical spin wave expansion, a fruitful direction is to compare a calculation of the lowest energy magnetic excitations of simple toy models with an exact solution for the spectrum of such toy models. A number of toy models exist for which an exact solution is available, at least in certain invariant subspaces. Examples include the Kondo dimer and the Kondo chain in one dimension in the extremely dilute limit. In cases where the exact ground state is a ferromagnet, a comparison of the lowest energy magnetic excitations can reveal how well the spin wave expansion approximates the exact energies. Such a comparison would furthermore provide key insight into the fundamental distinction between itinerant and Heisenberg ferromagnets, since in the latter the spin waves are exact eigenstates of the Hamiltonian, whereas they are not in Kondo lattice ferromagnets. A detailed investigation Kondo toy models will be presented elsewhere. 

\mbox{}

\section*{Acknowledgements}

We have greatly benefited from conversations and correspondence with Urban Seifert, Alexander Chernyshev, Rafael Fernandes, Pok Man Tam, Sanjeev Kumar, and Maria Daghofer. This research was supported by the National Science Foundation Award No. DMR-2144352.

\appendix

\section{Holstein-Primakoff expansion of $H_J$ \label{app:HP-J}}

This appendix provides a brief description of the HP expansion of the Heisenberg Hamiltonian $H_J$. Recall that in the local spin basis the Heisenberg Hamiltonian $H_J$ takes the form
\begin{multline}
H_J = \frac{1}{2S^2} \sum_{ij} J_{ij} \left\{ \Gamma^{zz}_{ij} S^z_i S^z_j + \left[ \frac12 A_{ij }S^-_iS^+_j + \right.\right. \\
\left.\left.  \frac12 B_{ij }S^-_iS^-_j + (\Gamma^{zx}_{ij} +i\Gamma^{zy}_{ij} )S^z_i S^-_j +\text{H.c.} \right] \right\} ,\label{app:H_J}
\end{multline}
see Eq.~\eqref{eq:H_J}, where $\Gamma^{ab}_{ij}  = \ehat^a_i\cdot \ehat^b_j $ were introduced in Eq.~\eqref{eq:Gamma} and the coefficients $A_{ij}$ and $B_{ij}$ are defined as
\beq
A_{ij } & =  &(\Gamma^{xx}_{ij} + \Gamma^{yy}_{ij}-i \Gamma^{xy}_{ij}+ i\Gamma^{yx}_{ij})/2, \\
B_{ij } & =  & (\Gamma^{xx}_{ij} - \Gamma^{yy}_{ij}+i \Gamma^{xy}_{ij}+i\Gamma^{yx}_{ij})/2.
\eeq

The $1/S$ HP expansion of Heisenberg models is a well-established standard technique~\cite{Holstein:1940p1098,Holstein:1941p388} and leads to a boson Hamiltonian of the general form
\be
H_J  = E_c + T^{(\frac12)}_{0 }+ T^{(1)}_{0 }+ T^{(\frac32)}_{0 }+\mathcal O(1/S^2). 
\ee
Here $E_c$ is the (classical) energy of the classical ordered ground state and the terms $T^{(q)}_{0 }$ collect contributions of order $S^{-q}$, where $q=n/2$ with $n=1,2,3,\dots$. Note that due to our definition of the exchange couplings (i.e., the overall prefactor of $1/S^2$ multiplying $H_J$) the classical energy is of order $S^0$. Note further that the notation of the  terms in the expansion is motivated by the structure and formalism of canonical spin wave expansion introduced in this work. That is to say, we group them into $T_{0 }$ since they manifestly do not constitute electronic spin-flip terms. 

The form of $E_c$ and $T^{(q)}_{0 }$ are found by substituting the HP boson representation of \eqref{eq:HP} and expanding in $1/S$.  The classical energy $E_c$ is given by
\be
E_c = \frac12 \sum_{ij}  J_{ij} \Gamma^{zz}_{ij},
\ee
and the operators $T^{(q)}_{0 }$ up to $q=2$ are found as
\begin{align}
T^{(\frac12)}_{0 } & = \frac{1}{\sqrt{2S}}\sum_{ij}J_{ij}\left[(\Gamma_{ij}^{zx}+i\Gamma_{ij}^{zy})a_j^{\dagger} + \text{H.c.} \right] \label{app:T_0^1/2} \\
T^{(1)}_{0 } &= \frac{1}{2S}\sum_{ij}J_{ij}\left(A_{ij}a_i^{\dagger}a_j + B_{ij}a_i^{\dagger}a_j^{\dagger} + \text{H.c.} \right) \nonumber \\
 &\hspace{2cm}     - \frac{1}{2S}\sum_{ij}J_{ij}\Gamma_{ij}^{zz}(a_i^{\dagger}a_i+a_j^{\dagger}a_j) \label{app:T_0^1} \\
    T^{(\frac32)}_{0 } & = \frac{-1}{S\sqrt{2S}}\sum_{ij}J_{ij}\Big[(\Gamma_{ij}^{zx}+i\Gamma_{ij}^{zy})(a_j^{\dagger}a_i^{\dagger}a_i + a_j^{\dagger}a_j^{\dagger}a_j)\nonumber \\
  &\hspace{5cm}    + \text{H.c.} \Big] \label{app:T_0^3/2} \\
       T^{(2)}_{0 } &= \frac{1}{2S^2}\sum_{ij}J_{ij}\Big[ \Gamma_{ij}^{zz}a_i^{\dagger}a_ia_j^{\dagger}a_j - \frac14 \Big(A_{ij}a_i^{\dagger}a_i^{\dagger}a_ia_j \nonumber \\ 
 &   + A_{ij}a_i^{\dagger}a_j^{\dagger}a_ja_j  + B_{ij}a_j^{\dagger}a_i^{\dagger}a_i^{\dagger}a_i +B_{ij} a_i^{\dagger}a_j^{\dagger}a_j^{\dagger}a_j \nonumber\\
 &  \hspace{5cm} + \text{H.c.} \Big)\Big] \label{app:T_0^2}
\end{align}
Notice that each boson operator comes with a $1/\sqrt{S}$ correction to the Heisenberg exchange coupling $J_{ij}$. In addition to these four terms, the classical energy of the Heisenberg model $\frac{1}{2}\sum_{ij}\Gamma_{ij}^{zz}J_{ij}$ comes from this procedure. This will be included in $T_0$ as well, but will not play a meaningful role in the commutators of the canonical transformation because it commutes with everything.



\section{Canonical transformation \label{app:canonical}}

In this appendix, we provide additional details about derivation of $\mathcal{H}^{(p)}$ and $Q^{(p)}$ outlined in Sec.~ \ref{ssec:derivation}. In particular, we show how to obtain $Q^{(2)}$, $\mathcal{H}^{(3)}$, and $Q^{(3)}$. Obtaining higher order terms in the series expansion is then straightforward. 

In Sec.~\ref{ssec:derivation} we determined $Q^{(1)}$ by requiring that all spin-flip terms at order $p=1$ are removed. Proceeding to order $p=2$, we collect all terms on the right hand side of Eq.~\eqref{eq:H-expand} and find
\begin{multline}
    [iQ^{(2)}, H_0] + [iQ^{(1)},T_0+T_1+T_{-1}]\\
    +\frac{1}{2}[iQ^{(1)},[iQ^{(1)},H_0]]
\end{multline}
Substituting $Q^{(1)}$ from Eq.~$\eqref{eq:Q1}$ and utilizing the commutation relation of Eq.~$\eqref{eq:commutator}$ yields
\begin{multline}
    [iQ^{(2)}, H_0] - \frac{1}{J_K}\Big([T_1,T_0] + [T_1,T_{-1}] + [T_0,T_{-1}]\Big)
\end{multline}
All terms without spin flips (satisfying $\sum_i m_i = 0$) are part of the Hamiltonian and we thus find $\mathcal{H}^{(2)} = -[T_1,T_{-1}]/J_K$, as quoted in Eq.~\eqref{eq:H(2)}. $Q^{(2)}$ is then determined by the requirement that it removes all spin-flip terms, which implies
\be
[iQ^{(2)}, H_0]= \frac{1}{J_K}[T_1 - T_{-1} , T_0].
\ee
Taking advantage of the commutation relation in Eq. $\eqref{eq:commutator}$, the form of $Q^{(2)}$ is straightforwardly found as
\be
iQ^{(2)} = \frac{1}{J_K^2}( [T_1 ,T_0] + [T_{-1}, T_0] ) \label{app:Q2}
\ee

With $Q^{(2)}$ and $\mathcal{H}^{(2)}$ fully determined, we proceed to order $p=3$. The sum of terms which contribute at this order are given by
\begin{multline}
    [iQ^{(3)},H_0] + [iQ^{(2)},T_0+T_1+T_{-1}] + \frac{1}{2}[iQ^{(1)},[iQ^{(2)},H_0]] \\
    + \frac{1}{2}[iQ^{(2)},[iQ^{(1)},H_0]] + \frac{1}{2}[iQ^{(1)},[iQ^{(1)},T_0+T_1+T_{-1}]] \\
    + \frac{1}{6}[iQ^{(1)},[iQ^{(1)},[iQ^{(1)},H_0]]]
\end{multline}
and inserting the previously found expressions for $Q^{(1)}$ and $Q^{(2)}$ gives, after simplifying,
\begin{multline}
    [iQ^{(3)},H_0] + \frac{1}{2J_K^2}\Big([[T_1,T_0],T_{-1}]+[[T_{-1},T_0],T_1]\Big) \\
    + \frac{1}{J_K^2}\Big([[T_1,T_0],T_0] + [[T_{-1},T_0],T_0]\Big) \\
    + \frac{1}{2J_K^2}\Big([[T_1,T_0],T_1] + [[T_{-1},T_0],T_{-1}]\Big) \\
    + \frac{2}{3J_K^2}\Big([T_1,[T_1,T_{-1}]]+[T_{-1},[T_{-1},T_1]]\Big) \label{app:p3}
\end{multline}
The terms which satisfy $\sum_i m_i = 0$ are part the Hamiltonian and form $\mathcal{H}^{(3)}$, whereas the remaining spin-flip terms determine $Q^{(3)}$. As seen in $\eqref{eq:H(3)}$, we find
\be
\mathcal{H}^{(3)} = \frac{1}{2J_K^2}\Big([[T_1,T_0],T_{-1}] + [[T_{-1},T_0],T_1]\Big),
\ee
and further find that $Q^{(3)} $ is determined by the equation
\begin{multline}
 [iQ^{(3)}, H_0]= - \frac{1}{J_K^2} \sum_{m=\pm1}\Big( [[T_m,T_0],T_0]  \\
+ \frac12 [[T_m,T_0],T_m] +\frac23 [T_m,[T_m,T_{- m}]]\Big),
\end{multline}
which may be inverted using the commutator $\eqref{eq:commutator}$; one finds
\begin{multline}
 iQ^{(3)}= - \frac{1}{J_K^3} \sum_{m=\pm1} m\Big( [[T_m,T_0],T_0]+ \frac14 [[T_m,T_0],T_m]  \\
 +\frac23 [T_m,[T_m,T_{- m}]]\Big). 
\end{multline}

Having $Q^{(3)}$ allows us to determine $\mathcal{H}^{(4)}$ which allows us to determine $Q^{(4)}$ etc. These steps can be continued until the desired order is reached. As mentioned previously, for the purposes of this work we have determined the Hamiltonian up to and including $\mathcal{H}^{(5)}$. For completeness we give the expressions here. We find
\begin{multline}
    \mathcal{H}^{(4)} = \frac{1}{2J_K^3}\sum_{m = \pm 1} m \Big( [T_m,[T_0,[T_{-m},[T_0]]]] \\
     - \frac{1}{2}[T_m,[T_{-m},[T_{-m},T_m]]] \Big),
\end{multline}
and 
\begin{multline}
    \mathcal{H}^{(5)} = \frac{1}{J_K^4}\sum_{m = \pm 1} \Big( \frac{7}{8}[T_m,[T_{-m},[T_m,[T_{-m},T_0]]]] \\
    - \frac{5}{8}[T_m,[T_m,[T_{-m},[T_{-m},T_0]]]] \\
    - \frac{1}{8}[T_m,[T_{-m},[T_{-m},[T_m,T_0]]]] \\
    + \frac{1}{2}[T_m,[T_0,[T_0,[T_{-m},T_0]]]]\Big).
\end{multline}
Depending on the desired orders in each of the perturbation parameters, one may need to determine additional terms in the Hamiltonian.


\section{Collecting Hamiltonian contributions from $\mathcal H^{(3)}$ \label{app:H(3)}}

Consider the third-order contribution to the Hamiltonian expansion in Eq.  \eqref{eq:H-expansion}, which takes the form \eqref{eq:H(3)} also shown here
\be
{\mathcal H}^{(3)} = \frac{1}{2J_K^2}\left([ [T_1,T_{0}],T_{-1}] +\text{H.c.}\right) \label{H(3)} 
\ee
Restricting the Hamiltonian expansion to linear spin wave order (1/S), the following contributions to the $T_1$ and $T_0$ operators are relevant ($T_{-1} = T_1^{\dagger}$)
\beq
T_1 &\simeq& T^{(t)}_1 +T^{(\frac12)}_1  \label{app:T_1}  \\
T_0 &\simeq& T^{(t)}_0 +T^{(S)}_0 + T^{(\frac12)}_0+ T^{(1)}_0, \label{app:T_0}
\eeq
where the forms in terms of the fermion and boson operators are given in Eq. \eqref{eq:T^t_1}, \eqref{eq:T^1/2_1} for $T_1$, and \eqref{eq:T_0^tS}, \eqref{app:T_0^1/2}, \eqref{app:T_0^1} for $T_0$. Up to linear spin wave order, the commutator $[[T_1,T_0],T_{-1}]$ has the following nine terms:
\begin{multline}
 [[T_1,T_0],T_{-1}] = [[T_1^{(t)},T_0^{(t)}],T_{-1}^{(t)}] + [[T_1^{(t)},T_0^{(t)}],T_{-1}^{(1/2)}] \\
+ [[T_1^{(1/2)},T_0^{(t)}],T_{-1}^{(t)}] + [[T_1^{(t)},T_0^{(S)}],T_{-1}^{(t)}] \\
+ [[T_1^{(1/2)},T_0^{(t)}],T_{-1}^{(1/2)}] + [[T_1^{(t)},T_0^{(1/2)}],T_{-1}^{(t)}] \\
+ [[T_1^{(t)},T_0^{(1)}],T_{-1}^{(t)}] + [[T_1^{(t)},T_0^{(1/2)}],T_{-1}^{(1/2)}] \\
+ [[T_1^{(1/2)},T_0^{(1/2)}],T_{-1}^{(t)}] \label{app:H3-expand}
\end{multline}
Each of these terms represent perturbative corrections to the hopping $t$ in the strong coupling $J_K \rightarrow \infty$ limit. The perturbations are in one or more of the small parameters $t/J_K$, $J/J_K$, $1/S$. Any term with a $1/S^{1/2}$ order will contribute to linear spin wave order in the second order diagrammatic calculation of the boson self-energy, as discussed in Sec. \ref{sec:SWT}. Since $T^{(t)} \sim t$ the first term is of order $t\left(t/J_K\right)^2$. This notation makes clear which perturbative corrections are present in each term. Since $T_1^{(1/2)} \sim 1/S^{1/2}$ the second and third terms are of order $t\left(t/J_K\right)\left(1/S^{1/2}\right)$. $T_0^{(S)} \sim 1/S$ so the fourth and fifth terms are of order $t\left(1/S\right)$. The sixth term has $T_0^{(1/2)} \sim J/S^{1/2}$ and is of order $t\left(t/J_K\right)\left(J/J_K\right)\left(1/S^{1/2}\right)$. Since $T_0^{(1)} \sim J/S$ the seventh term is of order $t\left(t/J_K\right)\left(J/J_K\right)\left(1/S\right)$. Finally, the eighth and ninth terms are of order $t\left(J/J_K\right)\left(1/S\right)$. Depending on the relative strength of $t$ and $J$, it may be more appropriate to consider an order like $tJ/\left(J_KS\right)$ to be a perturbative $t/J_K$ correction to the Heisenberg coupling $J$. In this case, we could rewrite the way we express the total order of terms six through nine. To illustrate this clearly, the order of the final term in \eqref{app:H3-expand} would be expressed as $J\left(t/J_K\right)\left(1/S\right)$.

The treatment of the commutator $[[T_1,T_0],T_{-1}]$ demonstrates how different contributions to the Hamiltonian can arise from $\mathcal{H}^{(3)}$. It also stresses the importance of collecting all terms of a given order correctly. Some of the orders present in $\mathcal{H}^{(3)}$ are also contained in $\mathcal{H}^{(4)}$. If one wishes to include quantum corrections to the linear spin wave order ($1/S^2$), additional terms would arise from the current expansion of $T_1$ in \eqref{app:T_1} and $T_0$ in \eqref{app:T_0}. However, to capture all possible contributions to this order on would also need to include $T_1^{(3/2)}$, $T_1^{(2)}$, $T_0^{(3/2)}$, and $T_0^{(2)}$.

\end{document}